%% file: main.tex
\documentclass[fleqn,usenatbib]{mnras}
\usepackage{newtxtext,newtxmath}
\usepackage[T1]{fontenc}
\usepackage{comment}
\DeclareRobustCommand{\VAN}[3]{#2}
\let\VANthebibliography\thebibliography
\def\thebibliography{\DeclareRobustCommand{\VAN}[3]{##3}\VANthebibliography}

\usepackage{graphicx}	% Including figure files
\usepackage{amsmath}% Advanced maths commands
\usepackage{amssymb}	% Extra maths symbols
\usepackage{soul}

\usepackage{tikz,xcolor,hyperref}
\definecolor{lime}{HTML}{A6CE39}
\DeclareRobustCommand{\orcidicon}{%
	\begin{tikzpicture}
	\draw[lime, fill=lime] (0,0) 
	circle [radius=0.16] 
	node[white] {{\fontfamily{qag}\selectfont \tiny ID}};
	\draw[white, fill=white] (-0.0625,0.095) 
	circle [radius=0.007];
	\end{tikzpicture}
	\hspace{-2mm}
}
\foreach \x in {A, ..., Z}{%
	\expandafter\xdef\csname orcid\x\endcsname{\noexpand\href{https://orcid.org/\csname orcidauthor\x\endcsname}{\noexpand\orcidicon}}
}

\title[Multi-wavelength study of 1ES 1218+304]{Multi-wavelength study of TeV blazar 1ES 1218+304 using gamma-ray, X-ray and optical observations }

\author[R. Diwan et al.]{
Rishank Diwan\orcidD$^{1}$\thanks{E-mail: rishank2610@gmail.com},
Raj Prince\orcidB$^{2}$,
Aditi Agarwal\orcidC$^{3,10}$,
Debanjan Bose\orcidA$^{4}$\thanks{E-mail: debaice@gmail.com},
Pratik Majumdar$^{5}$ ,\newauthor
Aykut \"Ozd\"onmez$^{6}$,
Sunil Chandra$^{7,8}$,
Rukaiya Khatoon$^{8}$,
Erg\"un Ege$^{9}$
\\
$^{1}$Laboratory for Space Research, The University of Hong Kong, 405B Cyberport 4, 100 Cyberport Road, Cyberport, Hong Kong\\
$^{2}$Center for Theoretical Physics, Polish Academy of Sciences, Al.\ Lotnik\'{o}w 32/46, 02-668 Warsaw, Poland\\
$^{3}$ Center for Cosmology and Science Popularization(CCSP) SGT University, Budhera, Delhi- NCR - 122006, India \\
$^{4}$ School of Astrophysics, Presidency University, Kolkata 700073, India\\
$^{5}$ Saha Institute of Nuclear Physics, a CI of Homi Bhabha National Institute, Kolkata 700064, West Bengal, India \\
$^{6}$ Ataturk University, Faculty of Science,  Department of Astronomy and Space Science, 25240, Yakutiye, Erzurum\\
$^{7}$ South African Astronomical Observatory, Observatory Road, Observatory, Cape Town 7925, South Africa  \\
$^{8}$ Center for Space Research, North-West University, Potchefstroom, 2520, South Africa \\
$^{9}$ Istanbul University, Faculty of Science, Department of Astronomy and Space Sciences, 34116, Beyazıt, Istanbul, Turkey \\
$^{10}$ Raman Research Institute, C. V. Raman Avenue, Sadashivanagar, Bengaluru - 560080, India
}

\date{Accepted 2023 July 6. Received 2023 July 6; in original form 2022 December 29}

\pubyear{2021}

\begin{document}
\label{firstpage}
\pagerange{\pageref{firstpage}--\pageref{lastpage}}
\maketitle

\begin{abstract}
We report the multi-wavelength study for a high-synchrotron-peaked BL Lac 1ES 1218+304 using near-simultaneous data obtained during the period from January 1, 2018, to May 31, 2021 (MJD 58119-59365) from various instruments including Fermi-LAT, Swift-XRT, AstroSat, and optical from Swift-UVOT $\&$ TUBITAK observatory in Turkey. The source was reported to be flaring in TeV $\gamma$-ray band during 2019, but no significant variation is observed with Fermi-LAT. A sub-hour variability is seen in the SXT light curve, suggesting a compact emission region for their variability. However, hour scale variability is observed in the $\gamma$-ray light curve. A "softer-when-brighter" trend is observed in $\gamma$-rays, and an opposite trend is seen in X-rays suggesting both emissions are produced via two different processes as expected from an HBL source. We have chosen the two epochs in January 2019 to study and compare their physical parameters. A joint fit of SXT and LAXPC provides  a constraint on the synchrotron peak, roughly estimated to be  $\sim$1.6 keV. A clear shift in the synchrotron peak is observed from  $\sim$1 keV to above 10 keV revealing its extreme nature or behaving like an EHBL-type source. The optical observation provides color-index variation as "blue-when-brighter". The broadband SED is fitted with a single-zone SSC model, and their parameters are discussed in the context of a TeV blazar and the possible mechanism behind the broadband emission.
\end{abstract}


%%%%%%%%%%%%%%%%%%%%%%%%%%%%%%%%%%%%%%%%%%%%%%%%%%

%%%%%%%%%%%%%%%%% BODY OF PAPER %%%%%%%%%%%%%%%%%%

\section{Introduction}
\label{intro}
Active galactic nuclei (AGN) host a supermassive black hole (SMBH) at the center, which accretes matter from the surrounding. The matters are in Keplerian orbit and fall into the SMBH via an accretion disk. The mechanism proposed in (\citet{1977MNRAS.179..433B}) suggests that the magnetic field lines from the accretion disk get twisted and collimated due to the high spin of SMBH and eject the matter through a bipolar jet perpendicular to the accretion disk plane. Later, the AGNs were classified based on how they are viewed, commonly known as the AGN unification scheme (\citealt{1995PASP..107..803U}). Blazars are a subclass of active galactic nuclei that have their relativistic jet pointed to the observer. They are characterized by rapid variability from hours to days timescales across all wavelengths, high polarization, and superluminal jet speeds. Blazars can be subdivided into flat spectrum radio quasars (FSRQs) and BL Lacertae (BL Lac) objects.
The broadband continuum spectra of blazars are dominated by non-thermal emission. The spectral energy distribution of blazars is characterized by a double hump structure: the first hump is generally attributed to the synchrotron radiation in the radio to X-ray bands, whereas there is intense debate about the origin of the second hump. The commonly accepted emission mechanism is via inverse Compton scattering of the low-energy photons by high-energy electrons in the system from GeV to TeV energies. There are alternative scenarios proposed by several authors which involve hadronic interactions producing neutral pions. These pions decay to generate photons in the GeV-TeV energies (\citealt{Mannheim1993, 2000NewA....5..377A, 2013ApJ...768...54B}). The BL Lac-type sources are further subdivided into three main classes depending on the position of their low-energy peak. If the synchrotron peak is observed at $<10^{14}$ Hz, those BL Lacs are called low-frequency peaked BL Lacs (LBLs). If the synchrotron peak is observed between $10^{14}$ Hz and $10^{15}$ Hz, they are called intermediate-frequency peaked BL Lacs (IBLs). Finally, BL Lacs with synchrotron peak $\geq 10^{15}$ Hz are called high-frequency peaked BL Lacs (HBLs). There is also a newly defined class of ultra-high-frequency peaked BL Lacs (UHBLs) with the spectral peak of the second bump (high energy peak) in the SED located at an energy of 1 TeV or above. These blazars are also known as "extreme blazars" or EHBLs. (\citealt{Abdo_2010}).

Multiwavelength observation of blazars is a very important tool for investigating the various properties of the blazars and the jet. For example, the shortest variability timescale allows one to put strong constraints on the size of the emission region of the blazar. The location of the emission region along the jet axis is another challenging problem in blazar physics. Many studies have been done in the past to locate the emission region; in some cases, it has been found that the emission happens very close to the SMBH within the broad-line region (BLR) (\citealt{Prince_2020, 2021MNRAS.502.5245P}). However, in some studies, it has been proposed to be at higher distances beyond the broad-line region (\citealt{10.1093/mnras/stt1723, Nalewajko_2014, 10.1093/mnras/stac1852}). The break or curvature in the $\gamma$-ray spectrum above 10-20 GeV suggests the emission region within the BLR as the BLR is opaque to high energy photons above 10 GeV (\citealt{Liu_2006}). The cross-correlation studies among the various wavebands are another way to locate the emission region along the jet axis. In many studies, it has been reported that simultaneous broadband emissions generally have a co-spatial origin. However, a significant time lag has been reported in some cases, strongly suggesting the different locations for the different emissions (\citealt{2019ApJ...871..101P}). In the first case scenario, one zone emission model is favored to explain the broadband SED, and in the later case, the multi-zone emission model is preferred (\citealt{2019ApJ...883..137P}).

The production of high-energy $\gamma$-rays in blazar suggests an acceleration of charged particles to very high energy, and many models have been proposed to explain the acceleration. The most accepted mechanisms are the diffusive shock acceleration (\citealt{1978MNRAS.182..147B}, \citealt{1978ApJ...221L..29B}, \citealt{1983RPPh...46..973D}, \citealt{1977DoSSR.234.1306K}, \citealt{1989ApJ...336..243S, 1989ApJ...336..264S}) and the magnetic reconnection (\citealt{2020NatCo..11.4176S}). In many studies, shock acceleration has been favored, which also demands the emission region close to the SMBH within the BLR because the shocks are produced and are strong at the base of the jet. On the other hand, the magnetic reconnection happens due to external perturbation and hence demands the jet to be less collimated, i.e., the emission region is farther from the base. \\

In this paper, we report on a multiwavelength study of the TeV blazar 1ES1218+304 to understand the broadband properties of the source. It is located at a redshift, z = 0.182 with R.A. = 12 21 26.3 (hh mm ss), Dec = +30 11 29 (dd mm ss). It has been observed in TeV energy with VERITAS (\citealt{2008AIPC.1085..565F}, \citealt{Veritas2009}) and MAGIC (\citealt{2006ApJ...642L.119A}, \citealt{MAGIC2011}) and is part of the TeV Catalog\footnote{\url{http://tevcat.uchicago.edu/}}.
%The paper is arranged in the following way. We discuss the multiwavelength observations and the data analysis procedures from different instruments used in this study in Section \ref{data_reduction}. In section \ref{results}, we have discussed the results from Astrosat alone and the broadband light curves and spectral energy distributions at length. In Section \ref{summary} we summarise and discussed the important findings in the context of blazar physics and eventually conclude our work in Section \ref{conclusion}. 

\section{Multiwavelength Observations, Data Analysis and Data Reduction} \label{data_reduction}
%The following section describes the data analysis technique used to generate a multi-waveband light curve. In the subsections, we provide a description of the data analysis technique of $\gamma$-ray data collected from {\it Fermi}-Lat. X-ray, and UV-optical data were collected from {\it Swift}-XRT and {\it Swift}-UVOT. Also, soft X-ray and hard X-ray data were collected from {\it AstroSat}-SXT and {\it AstroSat}-LAXPC, respectively and Optical Data from TUBITAK National Observatory.

\subsection{Fermi-LAT $\gamma$-ray Observatory} \label{Fermi_data}
Large Area Telescope (LAT) is a gamma-ray telescope placed on Fermi gamma-ray space observatory\footnote{\url{https://fermi.gsfc.nasa.gov/}} launched in 2008. It has a working energy range of 20 MeV to 1 TeV with a field of view of 2.4 Sr (\citealt{2009ApJ...697.1071A}). The orbital period of the telescope is around $\sim$ 96 mins in each hemisphere and covers the entire sky in a total $\sim$ 3 hr. Blazar 1ES 1218+304 is continuously being monitored since 2008. In this study, we have analyzed the data from 1st January 2018 - 31st May 2021, when the source was reported to be flaring in $\gamma$-rays (January 2019). The analysis was performed using \texttt{Fermipy v0.17.4}\footnote{\href{https://fermipy.readthedocs.io/en/latest/}{Fermipy webpage}}(\citealt{2017ICRC...35..824W}) and the standard Fermi tools software (Fermitools v1.2.23)\footnote{\href{https://github.com/fermi-lat/Fermitools-conda}{Fermtools Github page}} between 0.3-300 GeV. A 15$^\circ$ ROI was chosen around the source to extract the photon events with \texttt{evclass=128} and \texttt{evtype=3}, and the time intervals were restricted using ‘(DATA$\_$QUAL>0)\&\&(LAT$\_$CONFIG==1)’ as recommended by the Fermi-LAT team in the fermitools documentation. 
The source model file was generated using the Fermi 4FGL catalog \citep{Abdollahi_2020}, and the background $\gamma$-ray emission was taken care of by using the \texttt{gll\_iem\_V07.fits} file along with the isotropic background emission by using the \texttt{iso\_P8R3\_SOURCE\_V2\_v1.txt} file. In addition, the zenith angle cut was chosen as 90$^\circ$ to reduce the contamination from the Earth limb’s $\gamma$-ray. The source and background were modeled by the binned \texttt{Likelihood} method. Initially, the spectral parameters of all the sources along with those of galactic and isotropic background were kept free to optimize the $\gamma$-ray emission. Then to extract lightcurve and perform spectral fitting normalization of the sources only within 2$^\circ$ of ROI were kept free, and the rest of the parameters and other source models were frozen, except that of Source of Interest, in this case, blazar 1ES 1218+304 and a high flux source 4FGL J1217.9+3007, with an offset of 0.753$^\circ$ from 1ES 1218+304. Isotropic and Galactic background parameters were also kept free which in total constitutes to 10 parameters for likelihood analysis. PowerLaw model was used for the source as given below:
\begin{equation} \label{eq:1}
    \frac{d N (E)}{d E} = N_o \times \left( \frac{E}{E_o} \right)^{-\alpha}
\end{equation}
where E$_o$ and N$_o$ are the scale factor and the prefactor, respectively, provided in the 4FGL catalog, and $\alpha$ is the spectral index. Eventually, we generated the $\gamma$-ray light curves for 1, 3, 7, 14, and 30 days of binning for our scientific purpose.

\subsection{AstroSat}
On January 03, 2019, MAGIC reported a gamma-ray activity and detection of very high energy $\gamma$ -ray from the blazar 1ES 1218+304 (\citealt{2019ATel12354....1M}). Later, VERITAS also detected a $\gamma$-ray flare from this source (\citealt{2019ATel12360....1M}). Following these two events, we proposed a target of opportunity proposal in India’s first space-based multi-wavelength observatory, \texttt{AstroSat}\footnote{\href{https://www.isro.gov.in/AstroSat.html}{https://www.isro.gov.in/AstroSat.html}}. Observations were carried out from 17th to 20th January with a Soft X-ray telescope (SXT) and Large Area X-ray Proportional Counter (LAXPC).

\subsubsection{SXT}
The SXT working energy range is 0.3-7.0 keV, and the observation was performed with photon counting mode (PC). The level-1 data was downloaded from the webpage, and further reduction was performed with the latest SXT pipeline, \texttt{sxtpipeline1.4b} (Release Date: 2019-01-04). It produces the cleaned level-2 data products used for further analysis (\citealt{2016SPIE.9905E..1ES}, \citealt{2017JApA...38...29S}). The observations were done in various orbits; therefore, they were merged with the help of \texttt{SXTEVTMERGERTOOL}. The X-ray light curve is extracted using \texttt{XSELECT} with a circular region of 16$'$ centered on the source. The energy selection of 0.3-7.0 keV was applied in \texttt{XSELECT} using the channel filtering through the \texttt{pha\_cutoff} filter. The source spectrum was extracted for 0.3-7.0 keV energy range, and the background spectrum file used was provided by the \texttt{AstroSat} \texttt{SkyBkg$\_$comb$\_$EL3p5$\_$Cl$\_$Rd16p0$\_$v01.pha}. The spectrum was grouped in \texttt{GRPPHA} in order to have good photon statistics in each bin. The ancillary response file (arf) was generated using \texttt{sxtARFModule}, and the RMF file (sxt$\_$pc$\_$mat$\_$g0to12.rmf) was provided by the SXT-POC (Payload Operation Center) team. Eventually, the X-ray spectra from 0.3-7.0 keV with proper background and response files were loaded in \texttt{XSPEC} and fitted with the simple absorbed power-law and log-parabola spectral models. We use the galactic value, N$_{H}$ = 1.91$\times$10$^{20}$ cm$^{-2}$, for the correction of the ISM absorption model (\citealt{2016A&A...594A.116H}).
% A systematic uncertainty of 2$\%$ was added for the spectral analysis. %The fitting results can be seen in table \ref{table:3}.
%\textcolor{blue}{The merged cleaned event file then was used to select the source region of radius 12$'$ around the source and the background file was used provided by the \texttt{AstroSat} \texttt{SkyBkg$\_$comb$\_$EL3p5$\_$Cl$\_$Rd16p0$\_$v01.pha}} The tool \texttt{XSELECT} is used on the merged cleaned event file to select the source  region of radius 0.2$^\circ$ around the source The tool \texttt{XSELECT} was later used to extract the light curve and the spectrum. The spectrum was grouped in \texttt{GRPPHA} in order to have good photon statistics in each bin. The ancillary response file (arf) and RMF files were provided by the  SXT-POC (Payload Operation Center) team. Eventually, the X-ray spectra with proper background and response files were loaded in \texttt{XSPEC} and fitted with the simple absorbed power-law spectral model.

\input{Tables/Astrosat_SXT}
%\begin{figure}
%\begin{center}
%\includegraphics[height=\linewidth, angle = 270,trim={2.5cm 0.5cm 0.5cm 5cm},clip]{Figures/po_fix_nh.pdf}
%\caption{The 0.3 - 7.0 keV energy spectrum of 1ES 1218+304 from the data provided by {\it AstroSat}-SXT instrument during the period 17-20 January 2019.}
%\label{fig:sxt_spectrum}
%\end{center}
%\end{figure}

\subsubsection{LAXPC}
LAXPC works in the hard X-ray energy range from 3.0-80.0 keV (\citealt{2016SPIE.9905E..1DY}), consisting of three identical detectors namely LAXPC10, LAXPC20, and LAXPC30. Unfortunately, LAXPC10 was operating at a lower gain during the time of observation period. Also, the LAXPC30 detector has a gain instability issue caused by substantial gas leakage. Therefore, we used only LAXPC20 for the analysis, and the corresponding results are presented here. 

The Level-1 data were processed using the \texttt{LaxpcSoft} package available in AstroSat Science Support Cell (ASSC)\footnote{\href{http://astrosat-ssc.iucaa.in}{http://astrosat-ssc.iucaa.in}}. We generated the Level-2 combined event file using the command \texttt{laxpc\_make\_event}. During the data processing, a good time interval was applied to exclude the time intervals corresponding to the Earth occultation periods, SAA passage, and standard elevation angle screening criteria by using the \texttt{laxpc\_make\_stdgti} tool. Finally, the tools \texttt{laxpc\_make\_spectra} and \texttt{laxpc\_make\_lightcurve} were used to produce the spectra and lightcurve of the source using the \texttt{gti file}. We restricted the spectra to the 4-20 keV energy range since the background dominates the spectra above this energy. In the spectral analysis, 3\% systematic uncertainty was added to the data. The obtained lightcurve is not background subtracted, therefore, we estimated the background following the faint source routine \citep{2021JApA...42...55M}. However, due to insignificant variations observed in the extracted lightcurve from LAXPC20, we did not use them in our study.
%We created lightcurve and background lightcurve files using \texttt{laxpc\_make\_lightcurve} and \texttt{laxpc\_make\_backlightcurve} files for 3-20 keV for 1 sec binning, as LAXPC sees blazars as a faint source and therefore following the faint source pipeline, we limited our energy range to 20 keV. Due to insignificant variations observed in the extracted lightcurve from LAXPC20, we did not use them in our study.
 
%\begin{figure}
%\begin{center}
%\includegraphics[height=\linewidth, angle=270,trim={2cm 1cm 2cm 5cm},clip]{Figures/lc_500.pdf}
%\caption{500 Sec binned lightcurve from {\it AstroSat}-LAXPC instrument for 1ES 1218+304 during the period 17-20 January 2019.}
%\label{fig:laxpc_astrosat}
%\end{center}
%\end{figure}

\subsection{The Neil Gehrels Swift Observatory}
Simultaneous to AstroSat, blazar 1ES 1218+304 was also observed in X-rays with Swift-XRT and in optical-UV by Swift-UVOT telescopes\footnote{\url{https://swift.gsfc.nasa.gov/}}. %It provides a unique opportunity to have simultaneous broadband light curves and spectrum which is important to decipher the cause behind the flare and the broadband emission.

\subsubsection{XRT}
X-ray telescope (XRT) works in an energy range between 0.3-10.0 keV. Multiple observations were done during this period with an average of 2ks exposure. We have analyzed the data following the standard Swift \texttt{xrtpipeline}, and the details can be found on Swift webpage\footnote{\href{https://www.swift.ac.uk/analysis/xrt/}{https://www.swift.ac.uk/analysis/xrt/}}. The cleaned event files were produced, and a circular region of 10$"$ was chosen for the source and background around the source and away from the source. Tool \texttt{XSELECT} was used to extract the source light curve and the spectrum. The spectrum was binned using the tool \texttt{GRPPHA} to have a sufficient number of counts in each bin. A proper ancillary response file (ARF) and the redistribution matrix files (RMF) were used to model the X-ray spectra in \texttt{XSPEC}. A simple unabsorbed power law was used to fit the X-ray 0.3-10.0 keV spectra and extract the X-ray flux. The soft X-rays (below 1 keV) are affected by interstellar absorption in Milky-way and hence a correction is applied with N$_{H}$ = 1.91$\times$10$^{20}$ cm$^{-2}$ (\citealt{2016A&A...594A.116H}).

\subsubsection{UVOT}
Having an ultraviolet-optical telescope has the advantage of getting simultaneous observations to X-rays. UVOT has six filters: U, B, and V in optical and W1, M2, and W2 in the ultraviolet band. The image files were opened in DS9 software, and the source and background region of 5" and 10" were selected around the source and away from the source, respectively. The task \texttt{UVOTSOURCE} has been used to get the magnitudes, which were later corrected for galactic reddening, E(B-V)=0.0176 (\citealt{Schlafly_2011}) and converted into the fluxes using zero points and the conversion factor (\citealt{2006A&A...456..911G}).

\subsection{Optical}
The optical observations of our source were performed in the Johnson BVRI bands using the three ground-based facilities in Turkey, namely, 0.6m RC robotic (T60) and the 1.0m RC (T100) telescopes at TUBITAK National Observatory and 0.5m RC telescope at Ataturk University in Turkey. Technical details of these telescopes are explained in \citet{2022ApJ...933...42A}. The standard data reduction of all CCD frames, i.e., the bias subtraction, twilight flat-fielding, and cosmic-ray removal, was done as mentioned in \citep{2019MNRAS.488.4093A}.

\subsection{Archival}
We have used the archival optical data from ASAS-SN (All-Sky Automated Survey for Supernovae) (\citealt{2014ApJ...788...48S}; \citealt{2017PASP..129j4502K}) hown in lightcurve (Fig \ref{fig:lightcurve} last panel). We have also used long-term high flux observation in UV/Optical range from NASA/IPAC Extragalactic Database (NED)\footnote{\href{https://ned.ipac.caltech.edu/}{https://ned.ipac.caltech.edu/}} for providing the reference points in our SED analysis. We have also extracted the NuSTAR SED data points from (\citealt{Sahakyan2020}) and plotted them alongside our SED analysis.

\section{Results}
\label{results}
%In this section, we provide the main results of our work using the above broadband observations. We have explained various characteristics of broadband light curves and spectral energy distributions.

\subsection{Astrosat results} \label{astrosat_results}
\begin{figure}
    \centering
    \includegraphics[scale=0.36]{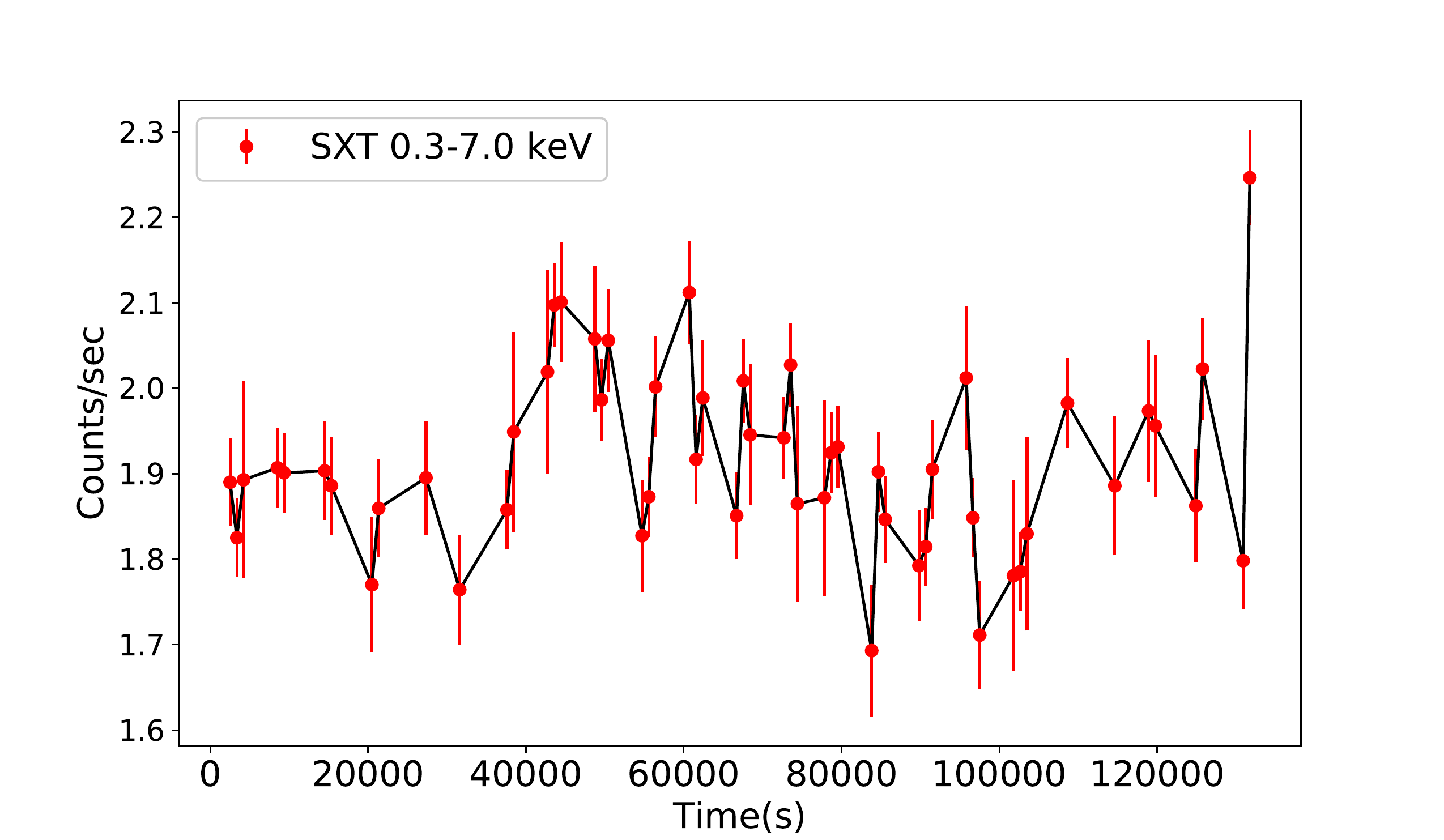}
    \caption{AstroSat-SXT light curve for energy 0.3-7.0 keV. The bin size is taken as 856 sec. }
    \label{fig:SXT-LC}
\end{figure}

\begin{figure}
\begin{center}
\includegraphics[scale=0.31]{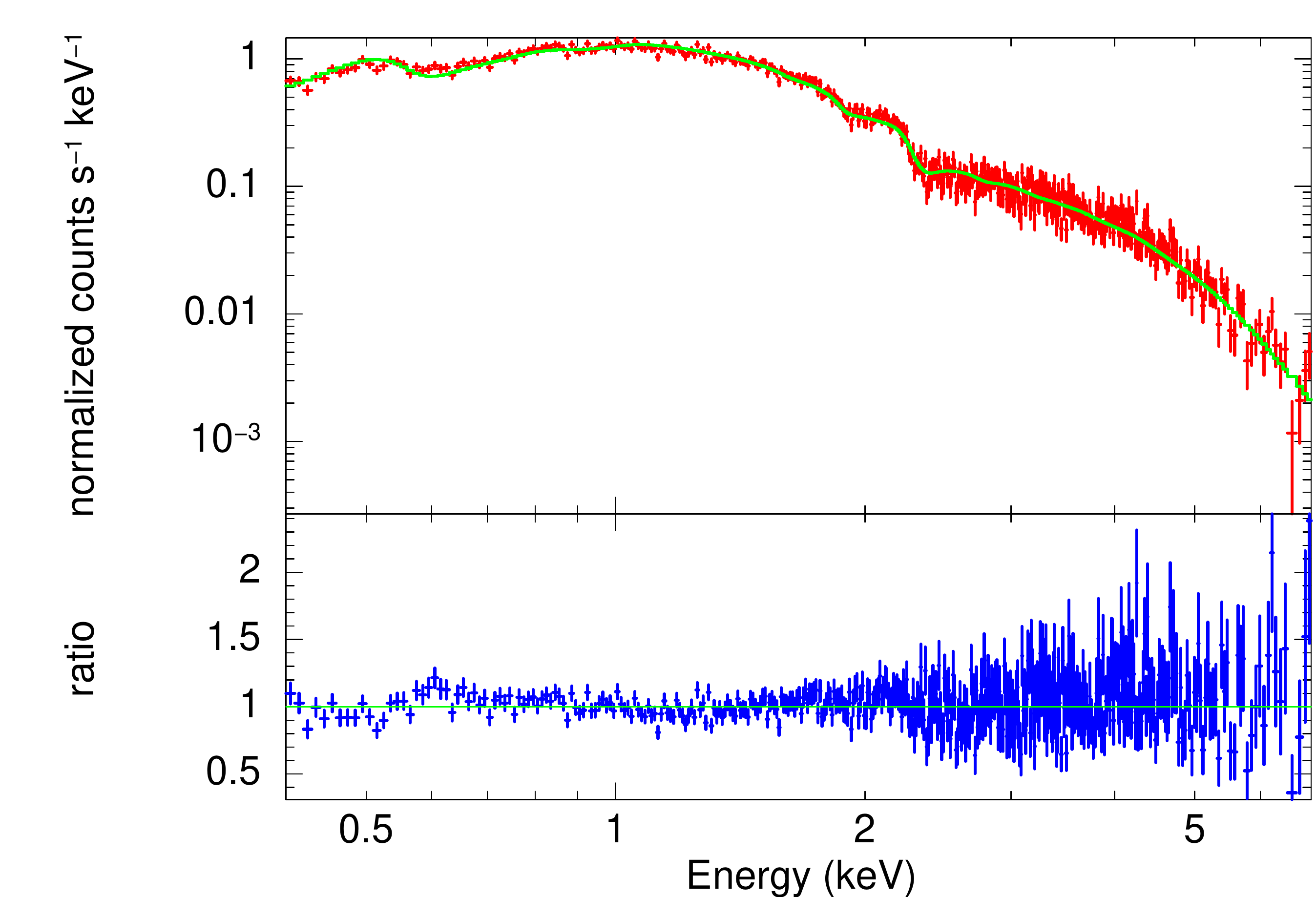}
\caption{Upper panel shows the 0.4 - 7.0 keV energy spectrum of 1ES 1218+304 fitted with Logparabola spectral model (green) with free galactic absorption. The SXT data (red) were taken during the period of 17-20 January 2019. Lower panel shows the ratio of data divided by folded model.}
\label{fig:sxt_spectrum}
\end{center}
\end{figure}

Astrosat observations in SXT and LAXPC were done during 17-20 January 2019 after two weeks of TeV detection. We have produced the SXT light curve and the spectrum as shown in Figure \ref{fig:SXT-LC} and Figure \ref{fig:sxt_spectrum} for the 0.3-7.0 keV energy band. The source appears to be variable on a short-time scale and the corresponding fractional variability and variability time is estimated in section \ref{blc}. A spectrum is extracted in the energy range of 0.3-7 keV and fitted with the power law and log-parabola models. The best-fit parameters are presented in Table \ref{table:3}. We started with a power-law model with fixed hydrogen column density, N$_H$ = 1.91$\times$10$^{20}$ cm$^{-2}$ %and ended up getting $\chi^2$/dof = 777/434 with photon spectral index, $\Gamma$ = 1.95$\pm$0.01 and 0.3-7 keV flux, F$_{0.3-7 keV}$ = (14.27$\pm$0.04)$\times$10$^{-11}$ ergs/cm$^2$/s. 
and next, we keep N$_H$ as a free parameter.%, and the best-fit value is estimated as 0.057$\pm$0.005 in units of 10$^{20}$ cm$^{-2}$. 
The fit gave the better-reduced chi-square in the latter case.
%The $\chi^2$/dof has improved to 595.75/433 and the spectral index was 2.11$\pm$0.02 with almost the same 0.3-7 keV flux. 
We repeat the same procedure with the log parabola model and with both fixed and free N$_H$, it gives a better fit than the power law. %With the free N$_H$ parameter we achieved a better fit with $\chi^2$/dof = 590.55/432 compared to the power-law case. The best-fit spectral index is 2.21$\pm$0.08 a bit softer than the power-law index. 
The details about the other parameters are provided in Table \ref{table:3}.

\begin{figure}
\begin{center}
\includegraphics[scale=0.31]{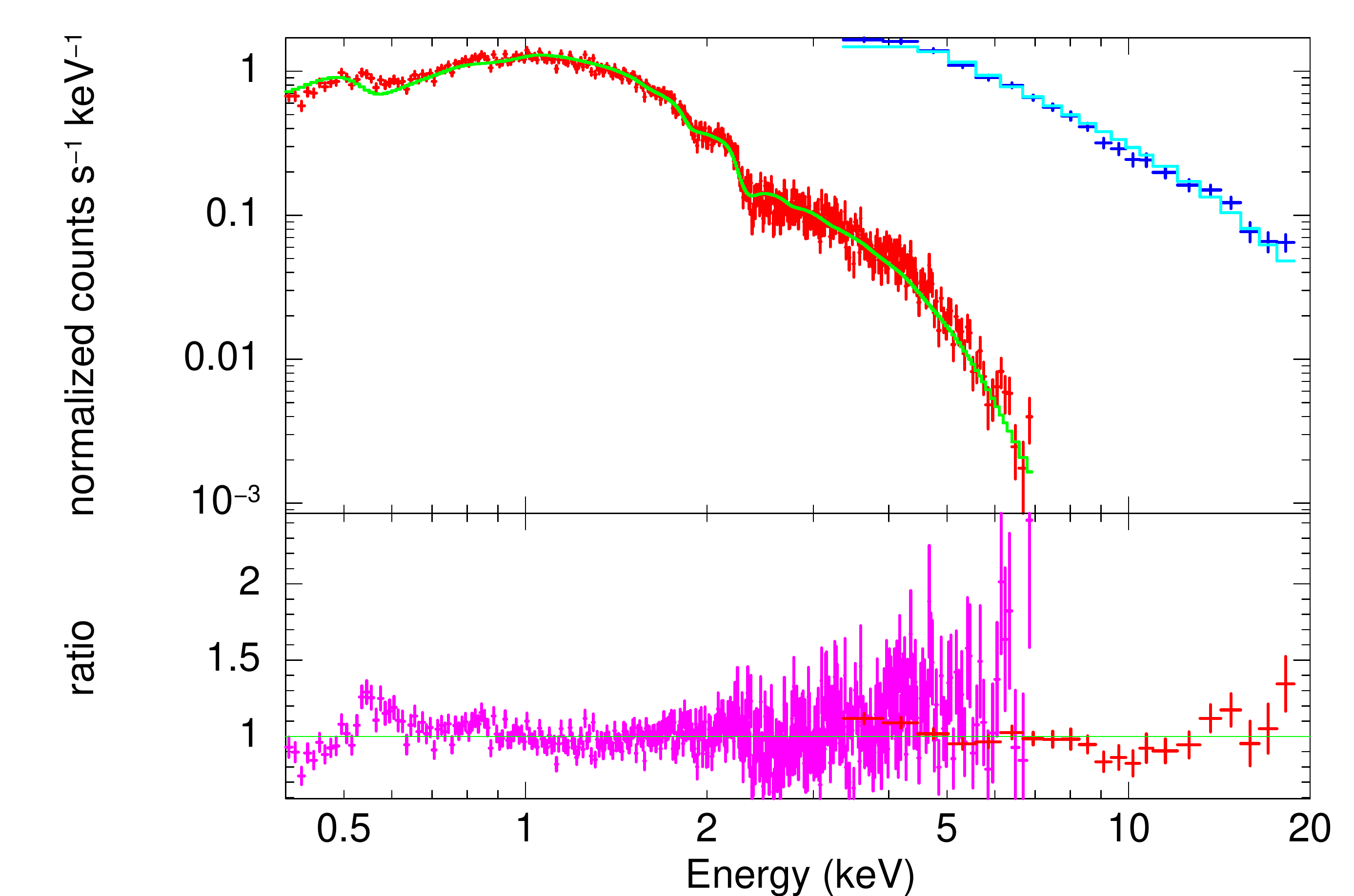}
\caption{In the upper panel the joint SXT (red) and LAXPC (blue) spectra are modeled together. The SXT energy range is taken as 0.3 - 7.0 keV and LAXPC is taken from 3.0-20.0 keV. The joint spectra are fitted with a log parabola spectral model (green + light blue). Both spectra were taken simultaneously during the period of 17-20 January 2019. The lower panel shows the ratio of data divided by folded model.}
\label{fig:joint-fit}
\end{center}
\end{figure}

We could not get a good light curve in LAXPC but extracted the spectrum from 4-20 keV. 
The simple power-law fit to LAXPC data alone gives a best fit power-law index, $\Gamma$ = 2.38$\pm$0.07 with fixed hydrogen column density.
The SXT and LAXPC spectra are jointly fitted with Power-law and Log-parabola models with free and fixed N$_H$ values. The log-parabola model provided a better fit to the spectrum (Table \ref{table:3}). 
%In the case of the Power-law, we get the $\chi^2$/dof = 948.46/403 and 623.25/402 for fixed and free N$_H$ values. 
%In both cases, the reduced-$\chi^2$ is much higher than the case of Log-parabola (Table \ref{table:3}) and hence not pursued further. 
For the joint fit, we used the total model as \texttt{constant*tbabs*logpar}. The constant factor is fixed at 1.0 for data group 1 and kept as a free parameter for data group 2. The best-fit value for the constant factor is 0.95$\pm$0.04 for both fixed and free N$_H$. The overall reduced-$\chi^2$ is improved when the N$_H$ is free and it is estimated as 4.2$\pm$1.0 ($\times$10$^{20}$ cm$^{-2}$).
%, almost two times higher than the fixed N$_H$ value. 
Figure \ref{fig:joint-fit} shows the best-fit plot with a log-parabola model. %We found that the spectral index, $\alpha$, and the curvature parameter, $\beta$, are slightly different during fixed and free N$_H$.
The mathematical representation of the log-parabolic model is given as,
\begin{equation}
    F(E) = K (E/E_1)^{(-\alpha+\beta log(E/E_1))} ph ~cm^{-2} ~s^{-1}~keV,
\end{equation}
where K is the normalization, and $E_1$ is the reference energy fixed at 1 keV. Using the best-fit parameters of the log-parabola model, we can estimate the location of the synchrotron peak, which is given as E$_p$ = E$_1$ 10$^{(2-\alpha)/2\beta}$ keV, where E$_1$ = 1.0 keV. For $\alpha$=1.88 and $\beta$=0.29, the E$_p$ is estimated as  1.6 keV or 3.9$\times$10$^{17}$ Hz. We also jointly fitted the SXT \& LAXPC with \texttt{eplogpar} model (total model as \texttt{constant*tbabs*eplogpar}) to estimate the location of the synchrotron peak and the estimated value is 1.54$\pm$0.34 keV which is consistent with E$_p$ estimated above.
%The peak of the synchrotron emission is mostly constrained by the X-rays, which peaks at $\sim$3.89$\times$10$^{17}$ Hz. %as shown in Figure \ref{fig:joint-fit}, 

\subsection{Broadband Light curves}
\label{blc}
\begin{figure*}
\begin{center}
\includegraphics[width=7in]{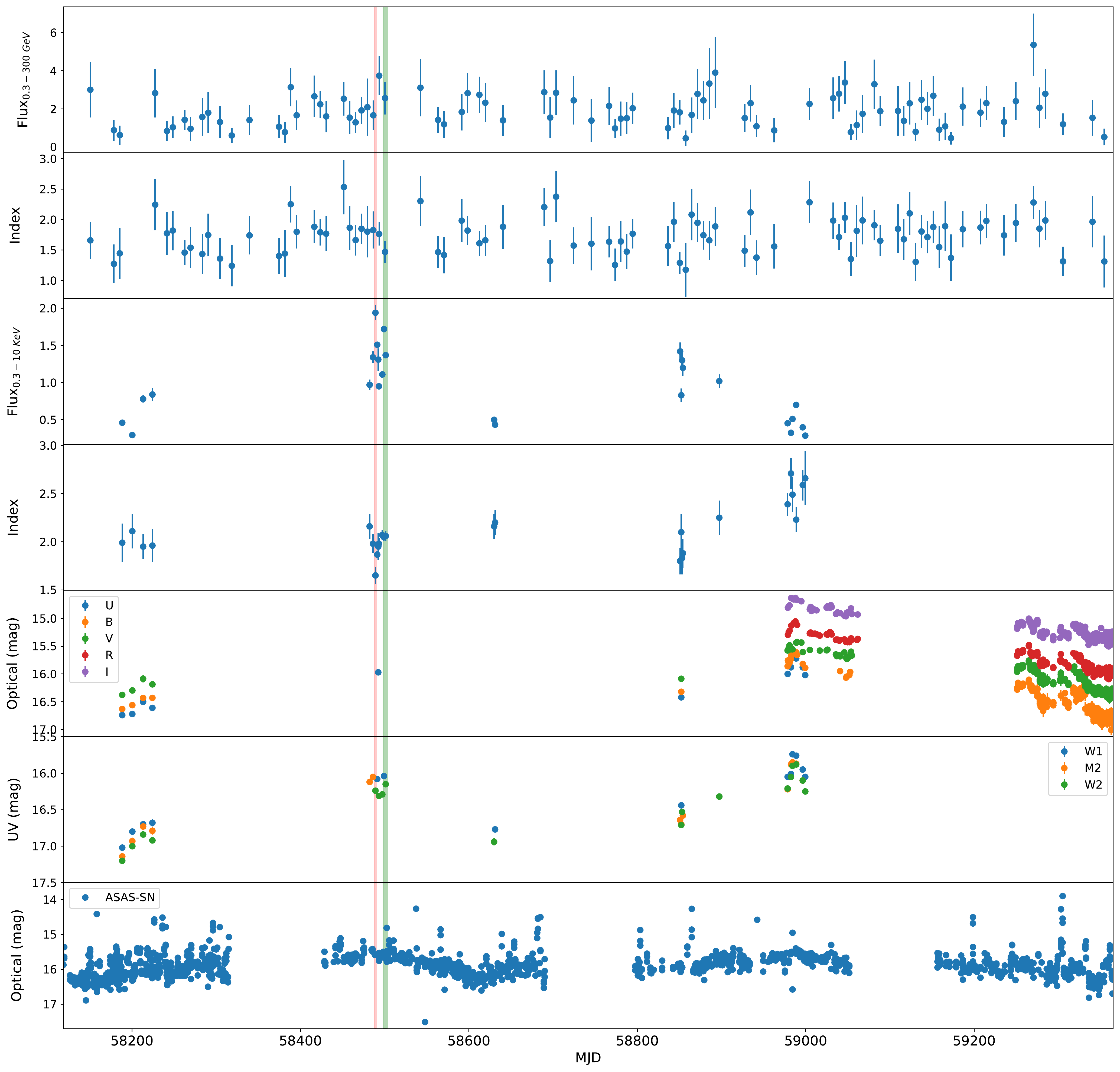}
\caption{Multi-wavelength light curve of 1ES 1218+304 from January 2018 to May 2021. 7-day binned $\gamma$-ray flux are presented in units of 10$^{-8}$ ph cm$^{-2}$ s$^{-1}$, and X-ray fluxes are in units of 10$^{-10}$ erg cm$^{-2}$ s$^{-1}$. The vertical red line represents the Flare period from 5-7 January 2019 and the vertical green line represents the Flare period from 15-20 January 2019. This period also includes the data from \texttt{AstroSat} for the period 17-20 January 2019. We identify these periods as Flare A and Flare B.}
\label{fig:lightcurve}
\end{center}
\end{figure*}

We have analyzed the $\gamma$-ray data between 2018 to 2021. the source was found to be in a flaring state in $\gamma$-ray band during Jan 2019. Furthermore, simultaneous observation in Swift-XRT and UVOT also confirms the flaring behavior in X-ray band as well as in optical-UV band. On 02 January 2019, the source was reported to be in a flaring state in very high energy $\gamma$-rays by MAGIC (\citealt{2019ATel12354....1M}), which was followed by VERITAS (\citealt{2019ATel12360....1M}). The observation done on 4, 5, and 6 January 2019 shows a high flux state above 100 GeV.  We identified this period as Flare A, marked by red color in Figure \ref{fig:lightcurve}. In X-rays and the optical range, the source was reported to be historically bright with flux around $\sim$2$\times$10$^{-10}$ erg cm$^{-2}$ s$^{-1}$ in X-ray band and with R band flux 2.35$\pm$0.05 mJy \citep{2019ATel12365....1R}. Based on this, we also proposed observation of this source using India's first space mission, AstroSat, for broadband observation. Our observation was done between 17-20 January 2019. This period is marked as a vertical green line in Figure \ref{fig:lightcurve} and identified as Flare B. The first two panels of Figure \ref{fig:lightcurve} represent the long-term $\gamma$-ray (GeV) light curve and corresponding photon spectral index. The source is not very bright in Fermi-LAT, but slight variability in the flux is seen. Panel 3 \& 4 represent the long-term Swift-XRT light curve and corresponding photon spectral index. A clear X-ray brightening is observed during Jan 2019. During this period, we do not have many optical observations (panel 5), and hence it's difficult to comment on the flux level. However, in UV (W1, M2, W2) bands (panel 6) high flux state is observed corresponding to TeV and X-ray activity. In panel 7, we show the archival optical data from ASAS-SN, and no short time scale variability is seen. We also have optical data from the ground-based observatory (panel 5), which covers the last part of the light curve showing a nice variation from a high flux state to a low flux state, suggesting a long-term variation in optical bands.

\subsection{Variability Study} \label{variability}
\begin{figure}
\begin{center}
\includegraphics[width=\linewidth]{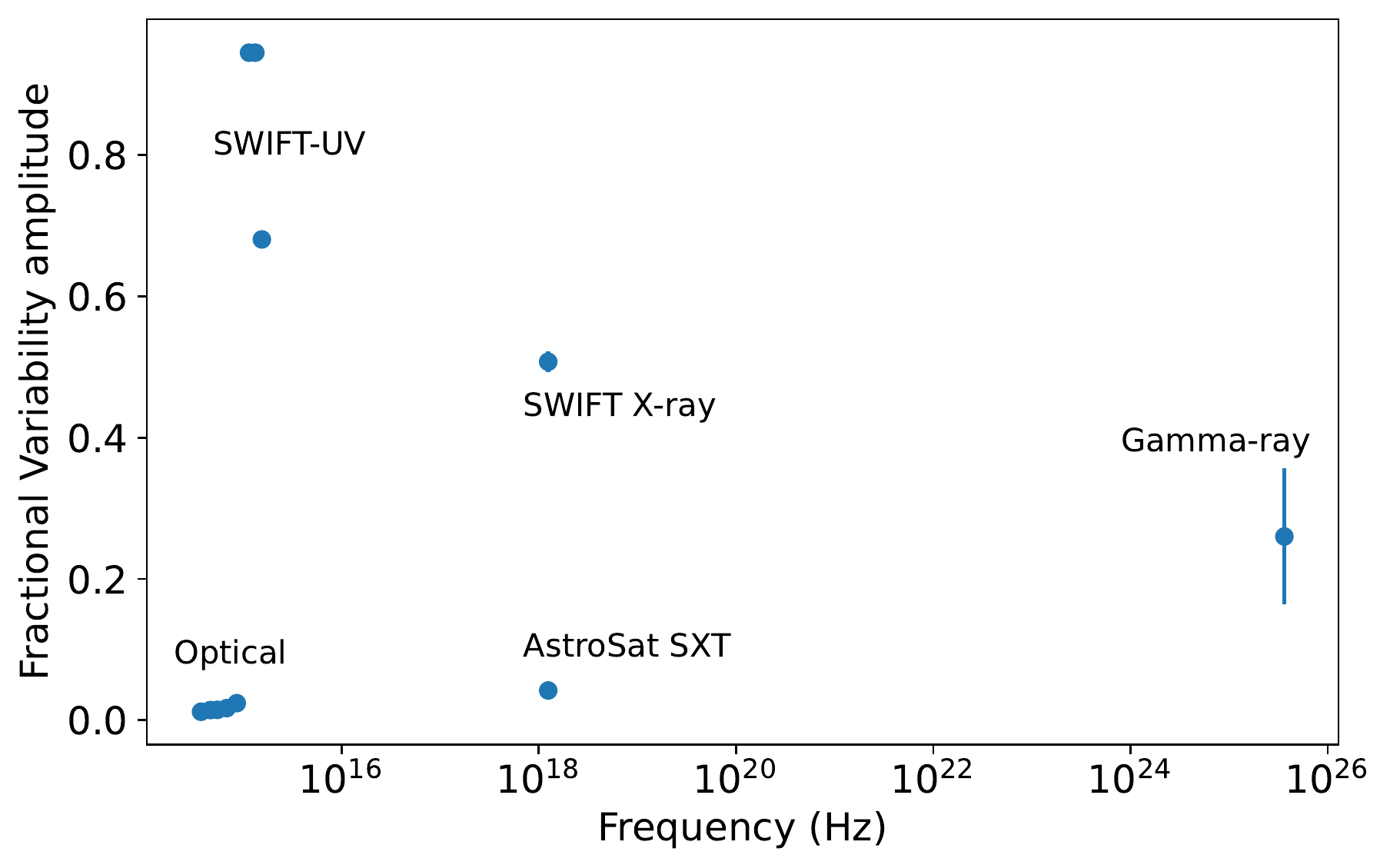}
\caption{Fractional variability for various wavebands is plotted with respect to their frequency.}
\label{fig:frac_var}
\end{center}
\end{figure}

%Fractional Variability
%{\bf \st{In general, blazar shows significant variability during the flaring period. The properties of these flares can depend on various factors like particle injection, particle acceleration, and energy dissipation in the jets of the blazars.}}
To study the intrinsic property, we calculate the Fractional Variability Amplitude (F$_{var}$) from the multi-wavelength light curve of the source. The relation given in (\citealt{2003MNRAS.345.1271V}) is used to determine the fractional variability (F$_{var}$)
\begin{equation}
    F_{var} = \sqrt{\frac{S^2 - E^2}{F^2}}
\end{equation}
\begin{equation}
    err(F_{var}) = \sqrt{\left( \sqrt{\frac{1}{2N}} \frac{E^2}{F^2 F_{var}}\right)^2 + \left( \sqrt{\frac{E^2}{N}} \frac{1}{F}\right)^2}
\end{equation}
where $S^2$ is the variance of the light curve, F is the average flux, $E^2$ is the mean of the squared error in the flux measurements, and N is the number of flux points in a light curve. We have estimated the F$_{var}$ for all the light curves, and the corresponding values are tabulated in Table \ref{table:6}. We observe that variability is more significantly detected in UV than X-rays and $\gamma$-rays. We also plot the F$_{var}$ with respect to the corresponding frequency in Figure \ref{fig:frac_var}. A similar behavior is also seen for another TeV blazar 1ES 1727+502 for one of the states \citep{10.1093/mnras/stac1866}. In past studies, it has also been argued that the variability pattern resembles the shape of the broadband SED seen in blazar if the source is observed from radio to very high energy $\gamma$-rays. One of the best examples is Mrk 421 which is also a TeV source, where the variability pattern during its two flaring states resembles the blazar SED \citep{2015A&A...576A.126A, 2015A&A...578A..22A}. A long-term study, using 10 yrs data, is done on 1ES 1218+304 by \citet{KKsingh_2019} using the multi-wavelength data from radio to $\gamma$-rays and the F$_{var}$ estimated on long-term period is different from what we have found in our study. \citet{KKsingh_2019} have found that source is more variable in radio at 15 GHz followed by X-ray band, optical-UV band, and $\gamma$-ray band.

\input{Tables/Fractional_variability.tex}

\par
% Calculate the values and write the details here.
%Time-Variability for Fermi
The timescale of variability is yet another important parameter that sets the bound on the size of the emission region. Doubling/Halving timescales are calculated from MJD 58119 to 59365 for the 7-day binned $\gamma$-ray light curve for all time bins. The formula used is:
\begin{equation}
    F(t_2) = F(t_1) \times 2^{(t_2 - t_1)/T_d}
\end{equation}
Here $F(t_1$) and $F(t_2$) are the fluxes measured at time $t_1$ and $t_2$, respectively. $T_d$ is the flux doubling/halving time scale, and also known as the variability time scale. The fastest variability time ($T_d$) in $\gamma$-rays was found to be 0.396 days or 9.5 hr. %The value for $t_{var}$ can be given by $t_{var} = ln(2) \times T_f$, which is 0.275 days or 6.6 hours. 
The hour scale variability is very common in blazar, suggesting a compact emitting region close to the central supermassive black hole.

%Time-Variability for Astrosat
Using the same equation, we also calculate the time-scale variability for the 856 sec binned AstroSat SXT light curve shown in Figure \ref{fig:SXT-LC}. The flux fastest variability time is estimated as  $T_d$ = 1849 sec. One can also assume that the fastest variability time can be linked with the exponential growth or decay time scale generally given by $T_f$ = $T_d$/ln2 which is estimated as  1281 sec (1.2 ksec) or $\sim$21 minutes. % $T_{var}$ = 2667.031 }. 
A similar flux variability time of 1.1 ksec is also estimated for Mrk 421 in the SXT light curve by \citet{2021arXiv210200919C}. Considering the fact that 1ES 1218+304 is a high synchrotron peaked blazar, the synchrotron process will produce the X-ray emission. We considered that the X-rays decay timescale corresponds to the radiation cooling timescale due to synchrotron emission \citep{2001ApJS..132..377H,2002ApJ...572..392B,2014A&ARv..22...72U}. Under this assumption, cooling time is equivalent to X-ray variability time, which is given by \citep{Rybicki1979},
\begin{equation}
    t_{cool} \simeq 7.74\times10^8  \frac{(1+z)}{\delta}B^{-2}\gamma^{-1} ~{\rm sec}.
\end{equation}
Where B is the strength of the magnetic field in Gauss and t$_{cool}$ is the synchrotron cooling timescale in seconds. Following \citep{Rybicki1979}, we can also derive the characteristic frequency of the electron population, 
responsible for the synchrotron emission at the SED peak,
\begin{equation} \label{eq:4}
    \nu_{ch,e} = 4.2\times10^6 \frac{\delta}{(1+z)}B \gamma^{2} ~{\rm Hz}.
\end{equation}
Using the above two equations, we eliminate the $\gamma$ since it changes with different states and derives a single equation given as,
\begin{equation}
    B^3\delta \simeq 2.5 (1+z)(\nu_{ch,e}/10^{18})^{-1} \tau_d^{-2}.
\end{equation}
Using the above equation, we derive the magnetic field strength for the Doppler factor, $\delta$ = 30, and variability time scale of 1.2 ksec, and it is found to be 0.1 G. The strength of the magnetic field derived from the broadband SED modeling is lower than the estimated value. This discrepancy could be because of the many assumptions made in deriving the equation (\ref{eq:4}) or due to the degeneracy in the SED modeling.

\subsection{Flux-Index Correlation}
\begin{figure}
\begin{center}
\includegraphics[width=\linewidth]{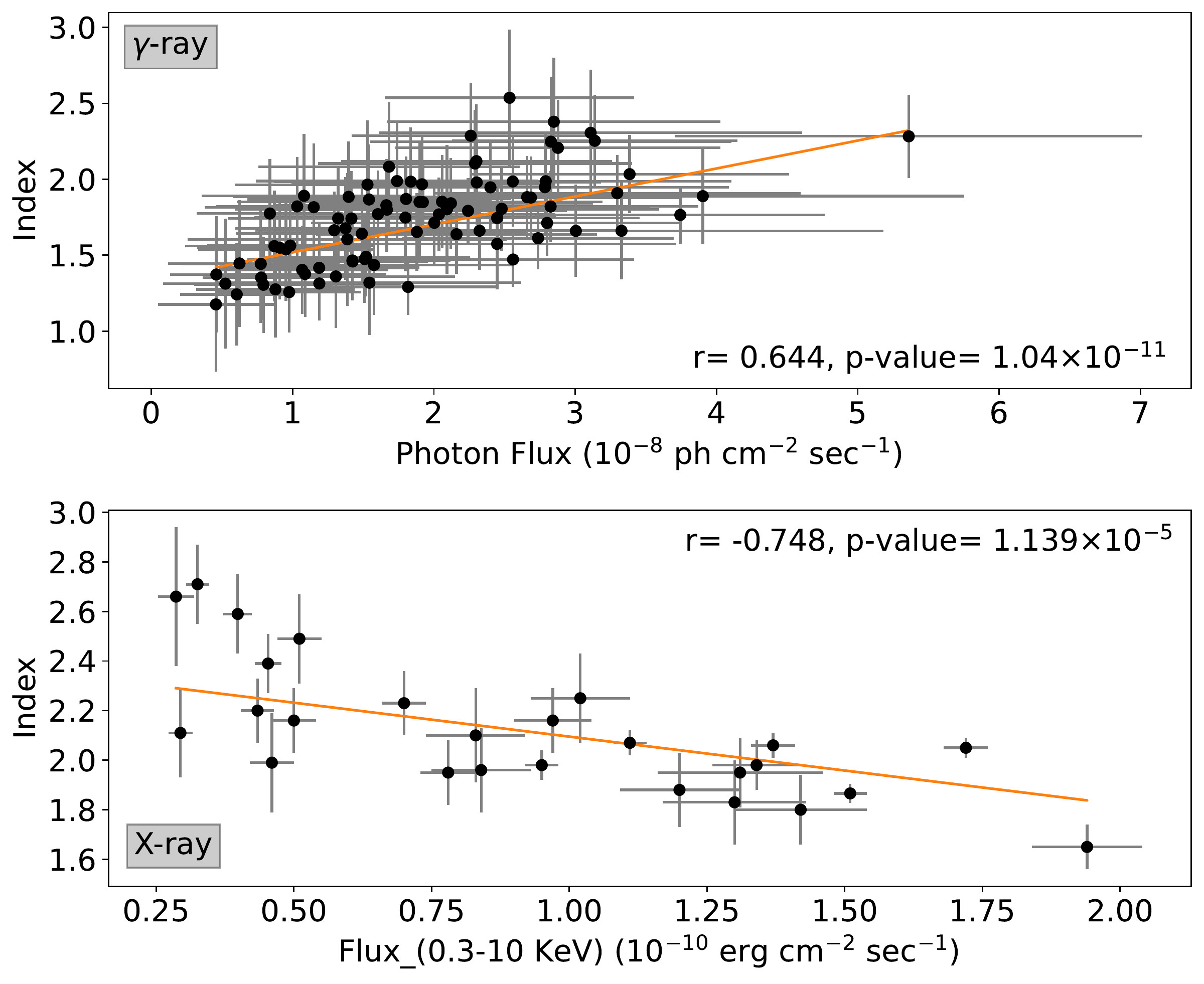}
\caption{Scatter plot for the correlation between flux and index of the blazar 1ES 1218+304. The top plot represents the 7-day binned Fermi-Lat data. The slope is positive and the Person correlation coefficient is 0.644. The bottom plot represents Swift-XRT data for Flux (0.3-10 keV) vs Photon Index. The slope is negative and the Pearson correlation coefficient is -0.748, it follows an inverse trend as the $\gamma$-ray data. The orange line is a linear fit for reference.}
\label{fig:Flux_index}
\end{center}
\end{figure}

We computed flux-index correlation for the $\gamma$-rays and X-rays data to study index hardening/softening. Figure \ref{fig:Flux_index} shows the flux vs. index plot with $\gamma$-ray band on the upper panel and X-ray band on the lower panel. In the case of $\gamma$-rays, we have taken data points with TS$\geq$16. We also observe a positive correlation between the flux and index, with Pearson correlation coefficient, R = 0.644 and p-value $\approx$ 0. The trend follows the linear function with slope = 0.184. In contrast to the above plot, X-rays data show an inverse trend, i.e., a negative correlation between flux and index, with Pearson correlation coefficient, R = -0.748 and p-value $\approx$ 0. It can also be fitted by a linear function with a slope = -0.274. This plot shows two contrasting trends, we can see the 'harder-when-brighter' trend in the X-ray energy range and the 'softer-when-brighter' trend in the $\gamma$-ray energy range. A similar trend is also observed for one of the TeV blazar 1ES 1727+502 \citep{10.1093/mnras/stac1866}. One of the possible explanations for having different trends in X-ray and $\gamma$-ray is that they are produced via two different processes. For BL Lac-type sources such as 1ES 1218+304, it is well-known that the synchrotron process produces the X-rays and $\gamma$-rays are produced via the inverse-Compton process.

A long-term study done by \citet{KKsingh_2019} also found a mild harder-when-brighter trend in X-rays using almost 10 yrs of data. The average spectral index is estimated as 1.99$\pm$0.16, which is consistent with our estimated value as $\sim$2.0 where synchrotron peaks in X-ray band. These results are also consistent with the long-term study done by \citet{10.1093/mnras/stw095} where they found the average photon spectral index as $\sim$2.0$\pm$0.01 for different values of galactic absorption taken from different models. A recent study by \citet{Sahakyan2020} estimated the average photon spectral index $\geq$2 for the period from 2008 to 2020. The spectra can be even harder during the bright state as 1.60$\pm$0.05, which is consistent with our result (see Figure \ref{fig:Flux_index}).

\subsection{Fermi-LAT $\gamma$-ray spectral fitting} \label{sed}
The data extraction and fitting process is provided in subsection \ref{Fermi_data}. We have used the \texttt{fermipy} to extract the $\gamma$-ray SED for the two periods (5-7 and 15-20 January 2019). The SEDs are then fitted with a simple power law spectral model. We noticed that the spectra are very hard and still increasing with energy suggesting the involvement of high-energy particles in their production. The fitted parameters are given in Table \ref{table:5} and the spectral index for period A ($\Gamma$=1.55$\pm$0.23) and B ($\Gamma$=1.54$\pm$0.19) is much harder than the average power law index, ($\Gamma$=1.75$\pm$0.03) for the total period. The harder spectra suggest that the IC peak is even at higher energy which is clearly seen in broadband SED modeling. A study by \citet{10.1093/mnras/sty857} also shows a harder $\gamma$-ray spectrum for many TeV blazar. A harder $\gamma$-ray spectrum is also seen in another TeV extreme blazar. Including the TeV data in broadband SED \citet{2022MNRAS.512.1557A} modeled the SED for six such sources with a two-zone emission model. Few new EHBL types sources are also discovered with the MAGIC telescope, and the Fermi-LAT $\gamma$-ray spectra were found to be very hard for all the sources suggesting an extreme location of the second SED peak above 100 GeV energy range \citep{2020ApJS..247...16A}. A long-term $\gamma$-ray spectral index was also estimated for 1ES 1218+304 by \citet{KKsingh_2019}, and they found it to be harder with 1.67$\pm$0.05, similar to our estimated value. \citet{Sahakyan2020} also estimated the $\gamma$-ray spectra averaged over $\sim$11.7 years which found to be 1.71$\pm$0.02 mostly consistent with above discussed results. These values are also consistent with the long-term average photon spectral index reported in the recent 4FGL catalog. 

\begin{figure}
\begin{center}
\includegraphics[width=\linewidth]{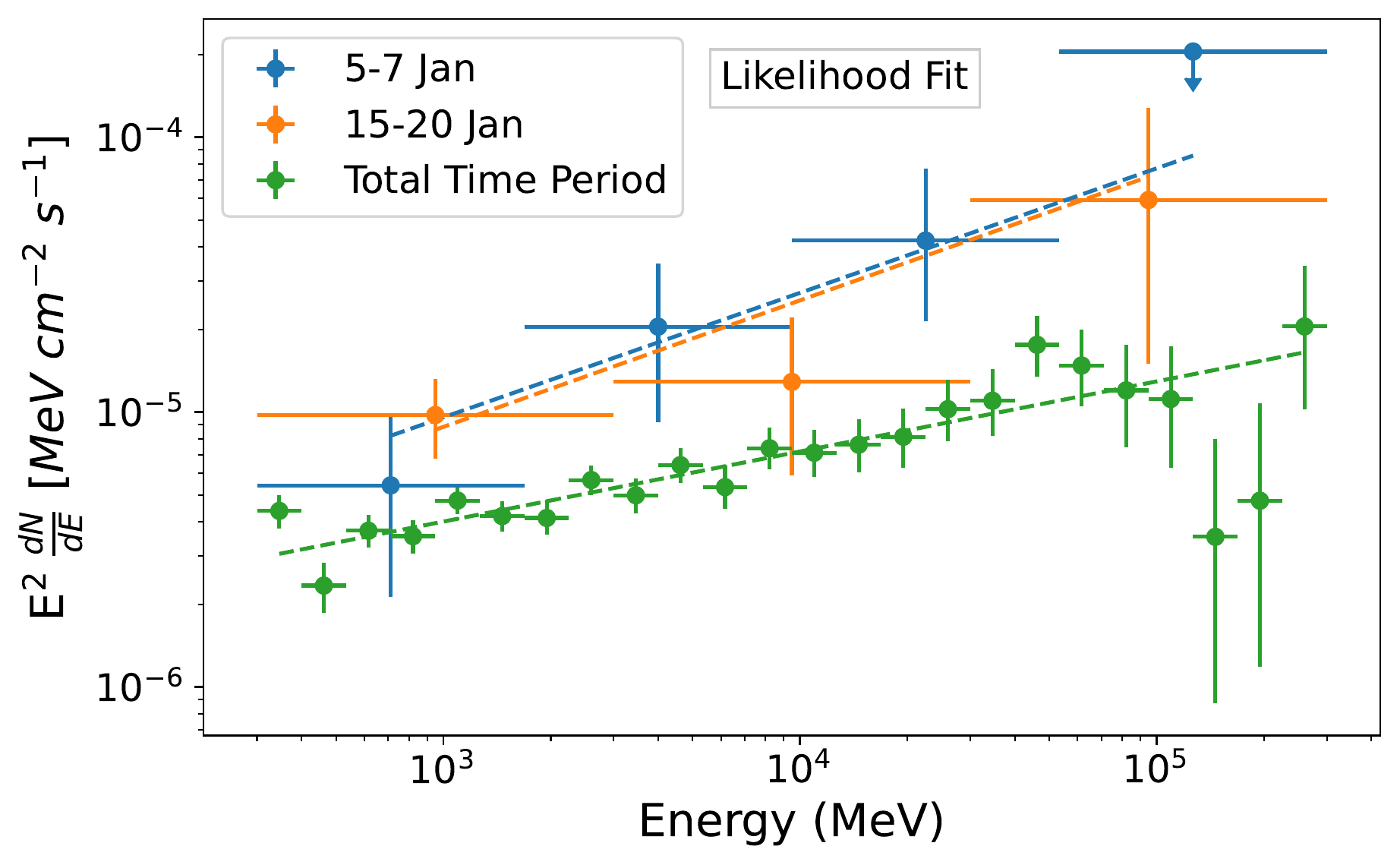}
\caption{The $\gamma$-ray SED extracted for both the period and fitted with power-law using the Likelihood fit method. The fitting parameters are discussed in the corresponding Section \ref{sed}.}
\label{fig:gamma_sed}
\end{center}
\end{figure}

\input{Tables/Fermi_spectrum_results.tex}
%\subsection{Correlation Study}: Since we studied only the correlation among different waveband emissions (no cross-correlation). Here we have to describe the 'r' and 'p' parameters of the correlation.
\subsection{Color-Magnitude Variations}
\begin{figure}
\begin{center}
\includegraphics[width=\linewidth]{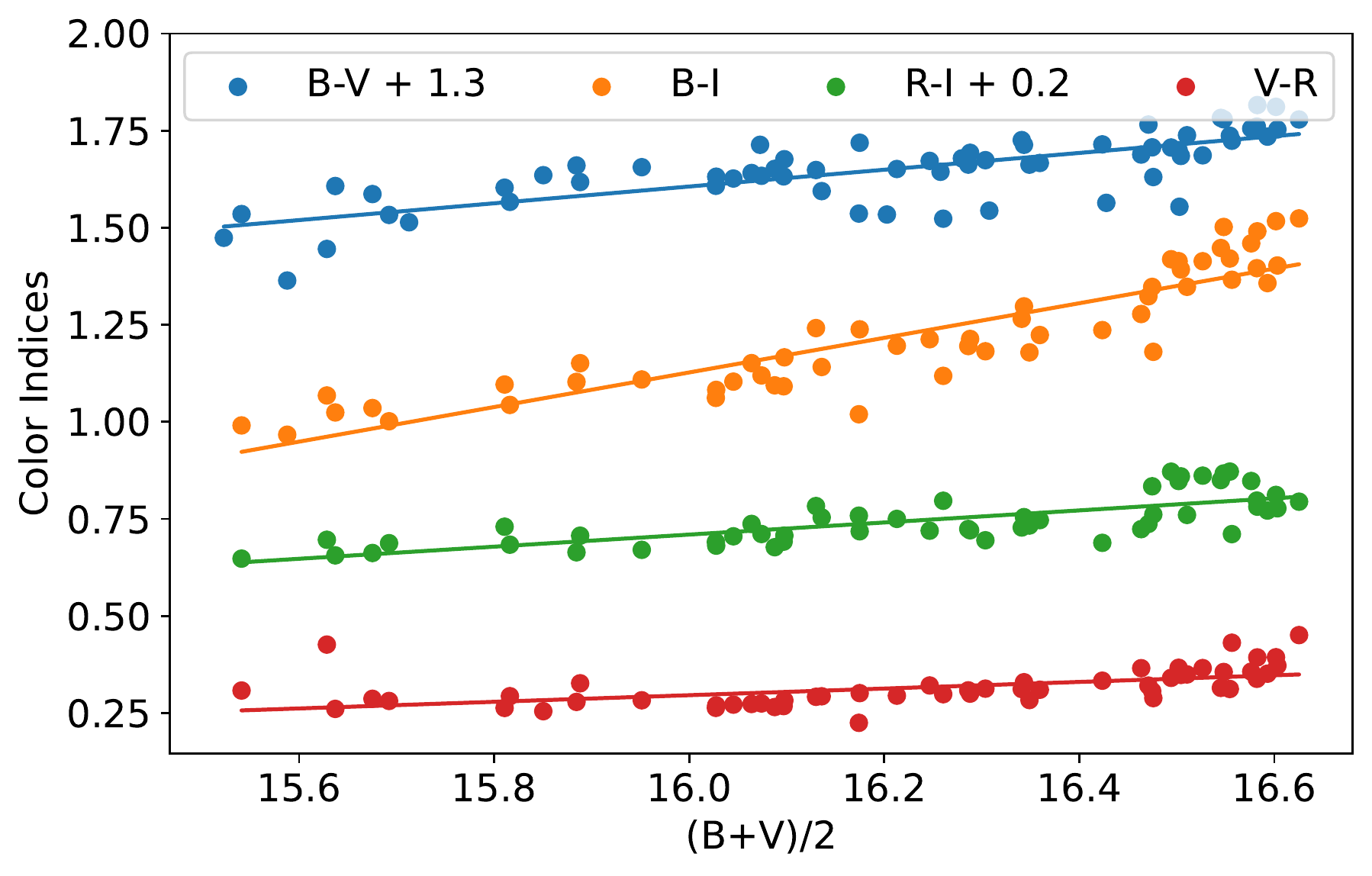}
\caption{Colour magnitude plot for 1ES 1218+304. The various color indices are plotted against (B+V)$/2$.}
\label{fig:Color_mag_diag}
\end{center}
\end{figure}

The color-magnitude relation helps us understand the different variability scenarios of the blazar. Fluctuations in optical flux are often followed by spectral changes. Therefore studying the color-magnitude (CM) relationship can further shed light on the dominant emission mechanisms in the blazar. To obtain a better understanding of the CM relation for our source, we fit a linear plot (CI = mV + c) between the color indices (CI) and (B+V)$/2$ magnitude. We then estimate the fit values, i.e., slope (m), constant (c), along with the correlation coefficient (r) and the respective null hypothesis probability (p) using two methods, Pearson and Spearman, as listed in Table \ref{table:1}. The generated CM plots are shown in Figure \ref{fig:Color_mag_diag}. Offsets of 1.3 and 0.2 are used for (B-V) and (R-I). A positive slope with p $<$ 0.05 implies a bluer-when-brighter (BWB) trend or a redder-when-fainter trend \citep{2021A&A...645A.137A} while a negative slope indicates a redder-when-brighter trend (RWB). As evident from Table \ref{table:1}, a significant BWB is dominant during our observation period for all possible color indices, namely; (B-V), (B-I), (R-I), and (V-R). Blazars, in general, display BWB from their quasi-simultaneous optical observations \citep{2000ApJS..127...11G, 2015MNRAS.451.3882A, 2016MNRAS.458.1127G}. 

The BWB trend can be attributed to the electron acceleration process to higher energies at the shock front, followed by losing energy by radiative cooling while propagating away \citep{1998A&A...333..452K}. On the other hand, the opposite trend of redder when brighter is observed more commonly in  FSRQs due to the contribution of bluer thermal emission from the accretion disc \citep{2006A&A...453..817V}. In addition to BWB and RWB trends, other optical studies have revealed cycle or loop-like trends \citep{2021A&A...645A.137A}, a mixed trend where BWB is dominant during higher state while RWB during the fainter state, or a stable-when-brighter (SWB) which is no significant color-magnitude correlation in the data at any timescale \citep{10.1093/mnras/stw377,2017ApJ...844..107I,2022MNRAS.510.1791N,2022ApJ...933...42A}. However, due to the lack of simultaneous observations for a larger sample of blazars, color-magnitude trends are still a topic of debate.

\input{Tables/Color_mag.tex}

\subsection{Broadband SED modeling}
\begin{figure}
\begin{center}
\includegraphics[width=\linewidth]{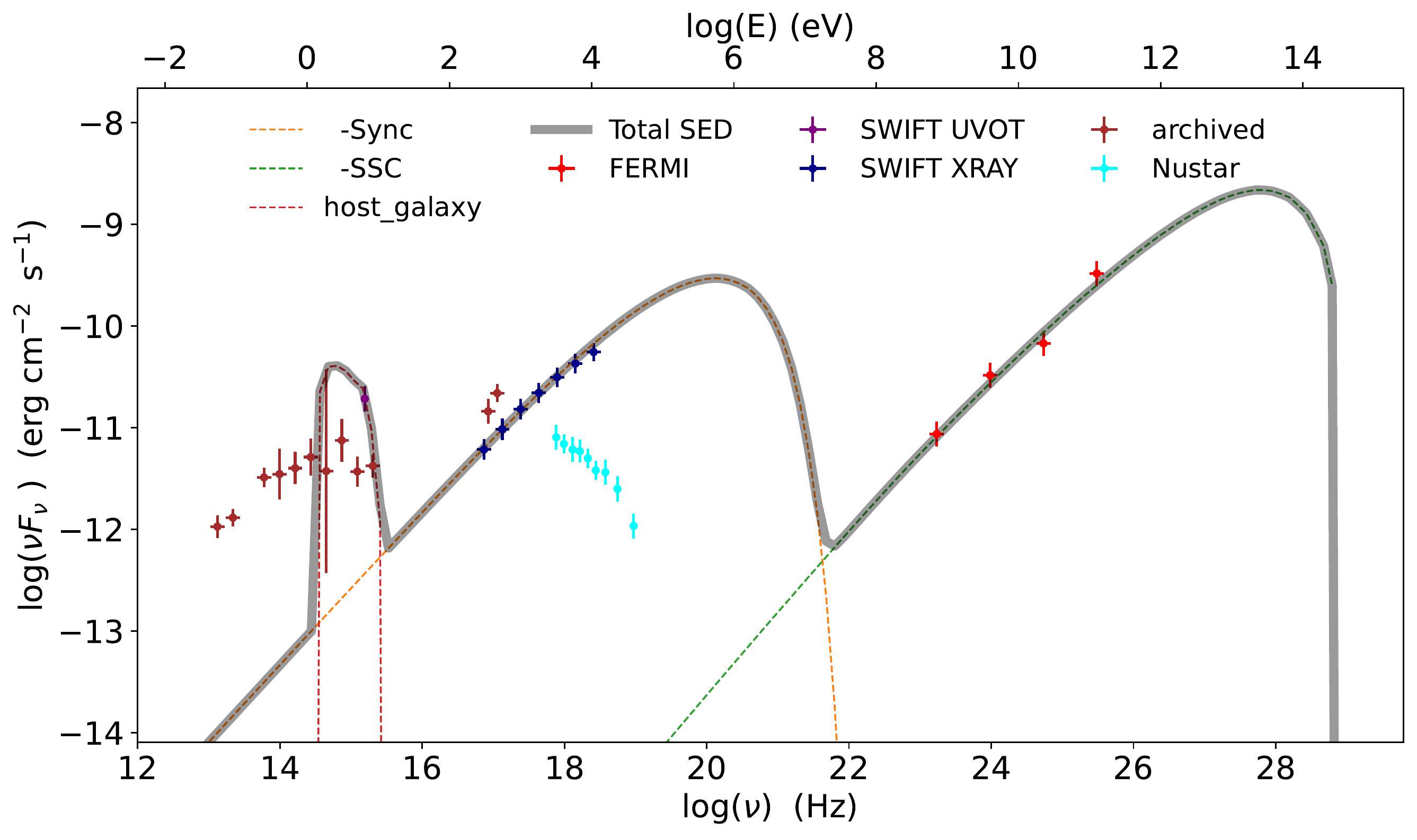}
\caption{Broadband SED Modelling for (5-7) January 2019 (Flare A). Optical data is fitted with the host galaxy template available in JetSet. Archival NuSTAR data are plotted in cyan color, which does not match the current X-ray spectral shape. Due to the hard X-ray spectral index,  the synchrotron peak is observed to be shifted to higher frequencies (above $\sim$10$^{18.5}$ Hz, possibly $\sim$10$^{20}$ Hz in our modeling) compared to the synchrotron peak location ($\sim$10$^{17.5}$ Hz) during (15-20) January as constrained by AstroSat observation in Figure \ref{fig:joint-fit} and also visible in Figure \ref{fig:15_20}.}
\label{fig:5_7}
\end{center}
\end{figure}

\begin{figure}
\begin{center}
\includegraphics[width=\linewidth]{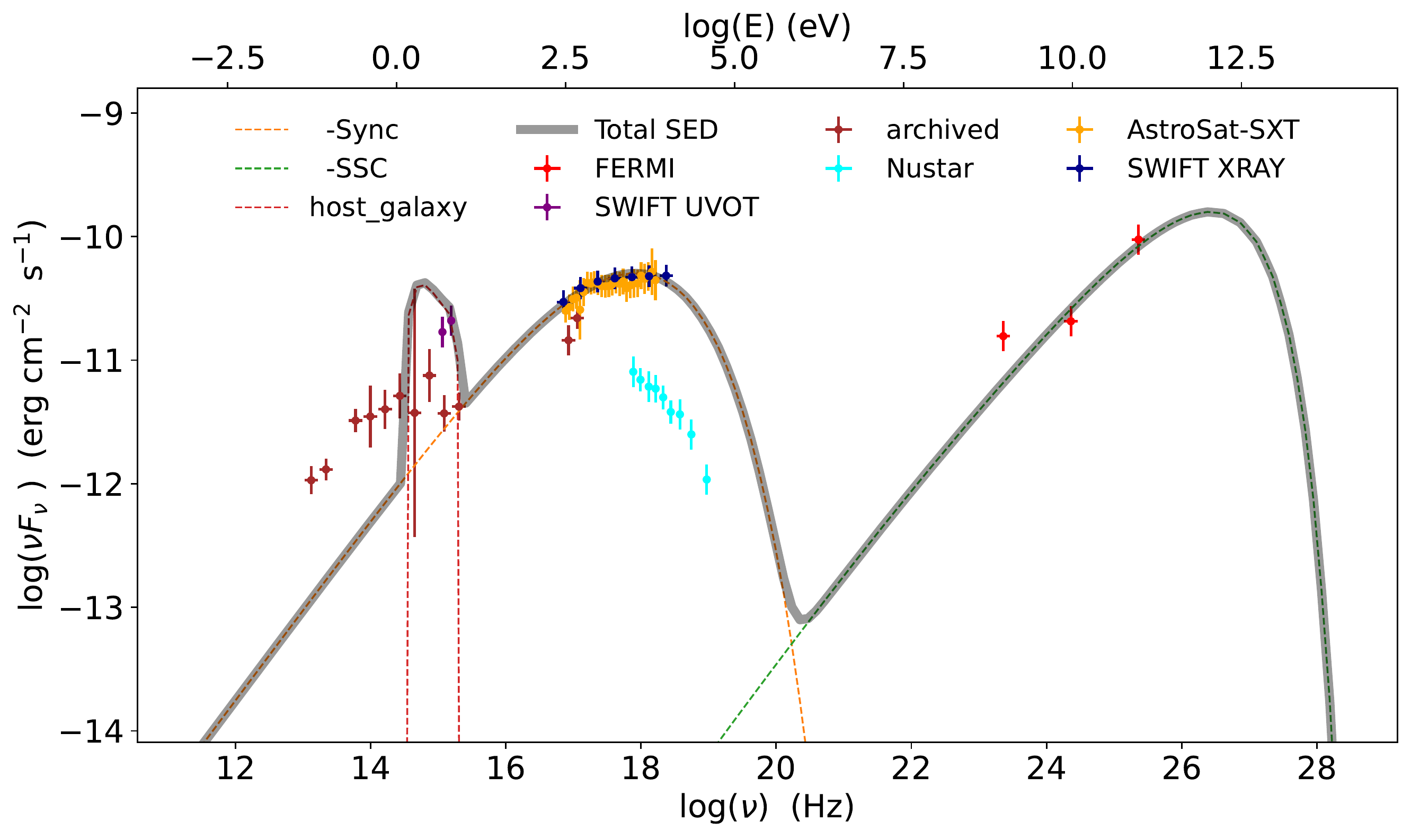}
\caption{The plot is the same as Figure \ref{fig:5_7} but for (15-20) January 2019 (Flare B). For this time period, the archival NuSTAR spectrum also does not match the current state X-ray spectral shape, suggesting that the NuSTAR spectrum was taken in low-flux states. The synchrotron peak is decided by the Swift-XRT and AstroSat-SXT spectra plotted on top of each other, which peaks at roughly $\sim$1.6 keV or 3.9$\times$10$^{17}$ Hz as estimated in section \ref{astrosat_results} using AstroSat data.}
\label{fig:15_20}
\end{center}
\end{figure}

%The broadband SED modeling in blazar is used to understand the simultaneous multi-wavelength emission from the source along with the possible physical mechanism responsible for broadband flaring event. 
Simultaneous multi-wavelength SEDs were generated for two time periods, which overlapped with the proposed flaring periods. The model fitting was done using a publicly available code JetSet\footnote{\href{https://jetset.readthedocs.io/en/latest/}{https://jetset.readthedocs.io/en/latest/}} (\citealt{2009A&A...501..879T}, \citeyear{2011ApJ...739...66T}, \citeyear{2020ascl.soft09001T}; \citealt{refId0}). Broadband emission of BL Lac sources like 1ES 1218+304 is better explained by the one-zone Synchrotron-Self Compton (SSC) model. Leptonic models assume that relativistic leptons (mostly electrons and positrons) interact with the magnetic field in the emission region and produce synchrotron photons in the frequency region of radio to soft X-rays or the first hump of the SED. The emission in the frequency region of X-rays to $\gamma$-rays or the second hump of the SED is produced by inverse Compton (IC) scattering of a photon population further classified into synchrotron-self Compton (SSC) or external Compton (EC) categories based on the source of the seed photons. In the case of SSC models (\citealt{GHISELLINI1993587}; \citealt{1992ApJ...397L...5M}), relativistic electrons up-scatter the same synchrotron photons which they have produced in the magnetic field. The model assumes a spherically symmetric blob of radius (R) in the emission region, surrounded by relativistic particles accelerated by the magnetic field (B). The blob makes an angle $\theta$ with the observer and moves along the jet with the bulk Lorentz factor $\Gamma$, affecting emission region by the beaming factor $\delta$ = $1/\Gamma(1-\beta \cos{\theta})$. The blob is filled with a relativistic population of electrons following an empirical Lepton distribution relation, and the power law with an exponential cut-off (PLEC) distribution of particles is assumed:
\begin{equation} \label{eq:2}
N_e(\gamma) = N_0 \gamma^{-\alpha} exp(-\gamma/\gamma_{cut})
\end{equation}
where $\gamma_{cut}$ is the highest energy cut-off in the electron spectrum. We see that the optical/UV measurements are higher than the non-thermal emission from the jet predicted by the SSC model. We also see high flux points in UV/optical range from the long-term observation of 1ES 1218+304, from NASA/IPAC Extragalactic Database (NED)\footnote{\href{https://ned.ipac.caltech.edu/}{https://ned.ipac.caltech.edu/}}. These observations suggest that the stellar emission from the host galaxy of the source is dominant at optical/UV frequencies. In order to accurately account for this emission due to the host galaxy, we have added the host galaxy component while modeling the SED using JetSet. Modeling of blazar 1ES 1218+304 is based on the SSC model in reference to equation (\ref{eq:2}). Results for the SSC model are shown in Figure \ref{fig:5_7} and Figure \ref{fig:15_20} for Flare A and Flare B. The model parameters are given in table \ref{table:4}.

\input{Tables/SED_model.tex}

\subsubsection{The constraint on Doppler factor}
We can calculate the minimum value of the Doppler factor using the detection of high-energy photons from the source. This calculation assumes the optical depth, $\tau_{\gamma \gamma}$(E$_h$), of the highest energy photon, E$_h$, to $\gamma \gamma$ interaction is 1. The target photons for optical depth of GeV $\gamma$-rays, caused by photon-photon collisions, are in the X-ray band. The formula for calculating the minimum value of the Doppler factor is derived in \citet{1995MNRAS.273..583D} \& \citet{Ackermann_2010, Ackermann2010} as
\begin{equation}
\delta_{min} \cong \left[ \frac{\sigma_t d^2_l (1+z)^2 f_{\epsilon} E_h}{4 t_{var} m_e c^4} \right] ^{1/6}
\end{equation}
where $\sigma_t$ is the Thomson scattering cross-section for the electron ($6.65 \times 10^{-25} cm^2$), d$_l$ is the luminosity distance of the source, $f_{\epsilon}$ is the X-ray flux in 0.3-10 keV energy range, E$_h$ is the highest energy photon, t$_{var}$ is the observed $\gamma$-rays variability time. For 1ES 1218+304, z=0.182, d$_l$ is 924 Mpc and t$_{var}$ is 0.396 days. Using the value of highest energy photon E$_h$ = 162.822 GeV for Flare A and 278.132 GeV for Flare B, and $f_{\epsilon}$ = $1.94\times 10^{-10}$ for Flare A and $1.55\times 10^{-10}$ for Flare B, we get the $\delta_{min}$ value to be 12.9 for Flare A and 13.6 for Flare B.

\subsubsection{The size of emission region}
%The information on the size and location of the emission region is very important for performing the SED modeling. 
%It gives us information on the jet axis, broadband emission, and what are the physical mechanisms involved. 
The variability time scale estimated from the $\gamma$-ray light curve is used to estimate the size of the emission region. The radius $R$ can be estimated by using the equation,
\begin{equation} \label{eq:3}
    R = c \delta t_{var}/(1+z),
\end{equation}
The Doppler factor from the broadband SED modeling is derived in \citet{KKsingh_2019} as $\delta$ = 26 and also assumed to be 25 in \citet{Sahakyan2020} which is higher than the minimum value derived in this paper. We took $\delta$ = 26 and the $R$ is estimated as $2.3 \times 10^{16}$ cm, using the and t$_{var}$ = 9.5 hr. During the SED modeling, 'R' is kept as free parameters, which gives the values $1.06\times10^{16}$ cm for Flare A and $1.40\times10^{16}$ cm for Flare B, both are consistent with the estimated value. The location of the emission region along the jet axis from the supermassive black hole can also be estimated from the variability time assuming a spherical emission region by using the expression $d \sim 2c \Gamma^2 t_{var}/(1 +z)$ \citep{Abdo_2011}. Using the conical jet geometry the Lorentz factor, $\Gamma$ = $\delta_{min}$, where the jet opening angle, $\theta_{jet}$ $<<$ 1/$\delta_{min}$ and $t_{var} = 0.396$ days and z = 0.182, the location is estimated to be, $d \sim 2.9 \times 10^{17}$ cm. To optimize the broadband SED modeling, we have fixed the location of the emission region to $1.0\times10^{17}$ cm from the SMBH along the jet axis.

\subsubsection{Jet Power}
We have estimated the power carried by individual components (leptons, protons, and magnetic fields) and the total jet power. The total power of the jet was estimated using
\begin{equation}
    P_{jet} = \pi R^2 \Gamma^2 c (U'_e + U'_p + U'_B)
\end{equation}
Here $\Gamma$ is the bulk Lorentz factor and $U'_e, U'_p, U'_B$ are the energy densities of electrons-positrons, cold protons, and the magnetic field respectively in the co-moving jet’s frame (primed quantities are in the co-moving jet frame while unprimed quantities are in the observer frame). All energy densities and the 'R' are the best-fit parameters of the fit and taken from Table \ref{table:4}. The power in leptons is given by
\begin{equation}
    P_e = \frac{3 \Gamma^2 c}{4 R} \int^{E_{min}}_{E_{max}} E Q(E) dE
\end{equation}
where Q(E) is the injected particle spectrum. The integration limits, $E_{min}$ and $E_{max}$ are calculated by multiplying the minimum and maximum Lorentz factor ($\gamma_{min}$ and $\gamma_{max}$) of the electrons with the rest-mass energy of the electron respectively. The power in the magnetic field is calculated using
\begin{equation}
    P_B = R^2 \Gamma^2 c \frac{B^2}{8}
\end{equation}
where B is the magnetic field strength obtained from the SED modeling. %The total jet power in protons is calculated by calculating the total energy led by protons and the volume of the emission region.
Our model returned the energy densities for electron-positron and magnetic field for both Flare events. The energy density for the cold proton was not estimated as it was too small. We calculated $P_e, P_B$, which is the power carried by the leptons and the magnetic field respectively. The total power $\mathrm{P_{jet} \approx P_{e} + P_{B}}$ along with the power of the individual components is shown in Table \ref{table:4}. The jet is dominated by the lepton's power compared to the magnetic field and can be considered a pair-dominated jet. The luminosities have been computed for a pure electron/positron jet since the proton content is not well known and can be considered as the lower limit. The absolute jet power $L_{jet} \simeq 1 \times 10^{46} erg s^{-1}$ for Flare A and is below the Eddington luminosity for a $5.6 \times 10^8 M_{\odot}$ black hole mass ($L_{Edd} = 7.3 \times 10^{46} erg s^{-1}$) estimated from the properties of the host galaxy in the optical band (\citealt{10.1111/j.1365-2966.2009.15738.x}). For Flare B, $L_{jet} \simeq 7.37 \times 10^{44} erg s^{-1}$ is significantly below the $L_{Edd}$.

\subsubsection{Broadband emission during flaring states}
We choose two flaring periods during the month of January 2019, MJD 58488-58490 (5-7 January 2019, Fig \ref{fig:5_7}) and MJD 58498-58503 (15-20 January 2019, Fig \ref{fig:15_20}) were modeled with a one-zone leptonic scenario. The modeled parameters are mentioned in Table \ref{table:4}. The model parameters inferred from this fitting suggest that Flare A had more activity compared to Flare B. The $\gamma_{max}$ and $\alpha$ are almost the same for both the flares, we see from Table \ref{table:4} that $\gamma_{min}$, $\gamma_{cut}$ have significantly higher values for Flare A compared to Flare B, which may be due to the flaring seen in the X-ray band. The magnetic field (B) for Flare A (2.74$\times$10$^{-3}$) is also less than that of Flare B (1.30$\times$10$^{-2}$). During the fitting of SED, we kept R$_H$ and $\delta$ as free parameters. We find that the value of R$_H$ is close to the value we calculate using equation (\ref{eq:3}). We also calculate the minimum doppler factor $\delta_{min}$ between the range (12.9-13.6), but during the SED modeling, we find that for Flare A, $\delta$ = 16 and for Flare B, it is much higher $\delta$ = 30 then the calculated value. It suggests that variation in $\delta$ could be one of the reasons for different flux states.

During these flares, the optical-UV emission is dominated by thermal emission from the host galaxy and hence has been modeled using the host galaxy model using JetSet. It is also seen that the X-ray data is better explained by synchrotron radiation of electrons. The SSC component of SED modeling dominates above 10$^{20}$ Hz ($\sim$ 1 MeV), and it is useful in describing the data up to the VHE $\gamma$-ray band.
%\subsubsection{Jet-Set}: about Jet-set: this we can also combined with above section or we can have separate here.

\section{Summary and Discussions}
\label{summary}
In our work, we present the multi-wavelength study of HBL blazar 1ES 1218+304 from 1st January 2018 to 31st March 2021 (58119-59365), which also includes the high flux event in VHE $\gamma$-rays detected by both MAGIC and VERITAS observatories during January 2019. This high flux rate was also seen in Swift-XRT and UVOT instruments. Hence we divided our SED analysis into two flaring periods, 5-7 January 2019 and 15-20 January 2019, for simultaneous multi-wavelength observation of 1ES 1218+304. The fastest variability timescale was found to be 0.396 days from analyzing the $\gamma$-ray light curve, constraining the size of the emission region to $\sim$$2.3 \times 10^{16}$ cm, which came out to be higher than previous modeling results (\citealt{10.1111/j.1365-2966.2009.15738.x}, \citealt{Sahakyan2020}, \citealt{KKsingh_2019}) but consistent with our SED modeling results, see Table \ref{table:4}. % and (\citealt{2021A&A...654A..96Z}) as during the SED modeling Doppler factor and the size of the emission region are optimized to best-fit value. 
The location of the emission region is estimated to be $d \sim2.9 \times 10^{17}$cm, which is similar to that used for SED modeling. The highest energy photon detected was 278.132 GeV which arrived during Flare B. We can also see the 'harder-when-brighter' trend in the X-ray energy range and the 'softer-when-brighter' trend in the $\gamma$-ray energy range. We also observed variability of 20 minutes in the SXT light curve. The SXT spectrum is well-fitted with the log-parabola model. The joint fit of SXT along with the LAXPC spectrum helps to constrain the location of the synchrotron peak. The estimated location is 1.6 keV which is roughly matching with the synchrotron peak constrained by the broadband SED modeling (Figure \ref{fig:15_20}).

The broadband SED modeling of the source was reproduced by a leptonic simple one-zone SSC model with the electron energy distribution described by a Power-law with an exponential cut-off (PLEC) function. Parameters like the magnetic field, injected electron spectrum, and minimum and maximum energy of injected electrons have been optimized to fit the SED's data points well. This study suggests that a single-zone model can also be good enough to explain the multi-waveband emissions from 1ES 1218+304 up to GeV energy range. The optical and UV emissions from the source are found to be dominated by the stellar thermal emissions from the host galaxy and can be modeled using the JetSet code by a simple blackbody approximation (\citealt{10.1111/j.1365-2966.2009.15738.x}).

\citet{10.1093/mnras/sty857} argued that the one-zone SSC model can explain the broadband SED modeling in hard-TeV blazar at the expense of extreme electron energies with very low radiative efficiency. The maximum electron Lorentz factor estimated in their modeling for all the six sources is orders of 10$^7$ which is consistent with our results for 1ES 1218+304. The other modeling parameters, such as the size of the emission region, magnetic field strength, and the magnetization parameters (U$_B$/U$_e$), are very similar to our SED modeling result for 1ES 1218+304. In our case, the U$_B$/U$_e$ = 10$^{-4}$ - 10$^{-6}$ and in \citet{10.1093/mnras/sty857} it is order of 10$^{-2}$ - 10$^{-5}$. Similar results were also obtained by \citet{2011A&A...534A.130K} where they model the broadband SED of extreme TeV source 1ES 0229+200. The magnetic field and the magnetization parameter (10$^{-5}$) are consistent with our results for 1ES 1218+304. However, their model requires a narrow electron energy distribution with $\gamma_{min}$ $\sim$ 10$^5$ to $\gamma_{max}$ $\sim$ 10$^7$ rather than the broad energy range obtained in our study, \citet{10.1093/mnras/sty857}, and \citet{2020ApJS..247...16A}.

\citet{2020ApJS..247...16A} have observed ten new TeV sources with MAGIC from 2010 to 2017 for a total period of 262 hours, and the simultaneous X-ray observations confirm that 8 out of 10 sources are of extreme nature. Their $\gamma$-SED was found to be very hard between 1.4 to 1.9. Blazar 1ES 1218+304 is also an extreme TeV blazar, and in our study, the $\gamma$-ray SED is found to be 1.5, consistent with the above TeV sources. They have modeled all the sources with a single zone conical-jet SSC model. Additionally, they also used the proton-synchrotron and a leptonic scenario with a structured jet. They also argue that all the model provides a good fit to the broadband SED, but the individual parameters in each model differ substantially. Comparing their SSC model results to our SSC modeling, the maximum electron energy is consistent. The electron spectral index, in our case, is harder than their results, and the magnetic field is much smaller. The estimated Lorentz factor is more or less consistent with the $\Gamma$ used for all the sources in their study. In their recent work, \citet{2022MNRAS.512.1557A} have modeled the six well-known extreme BL Lac sources with a lepto-hadronic two-zone emission model to explain the broadband SED. In another study, \citet{2021A&A...654A..96Z} have shown that the broadband SED of extreme BL Lac sources can be explained by considering the co-acceleration of electrons and protons on internal or recollimation shocks inside the relativistic jet.

\citet{Sahakyan2020} has modeled the average state of 1ES 1218+304 with one-zone SSC model. The parameter estimated in their study is mostly consistent with ours. However, our study focuses on the smaller period, including two flaring events. During the flaring event (15-20 Jan), the magnetic field and the magnetization parameters are estimated as 1.30$\times$10$^{-2}$ Gauss and $\sim$10$^{-4}$, which is comparable to the value for the same parameters estimated by modeling the average state of the source in \citet{Sahakyan2020}. However, the Doppler factor required in \citet{Sahakyan2020} is much higher than the Doppler factor needed to fit the flaring state in our case. \citet{KKsingh_2019} also modeled the average broadband SED collected for almost 10 years with a one-zone SSC model. The required $\gamma_{min}$, $\gamma_{max}$, and Doppler factor are consistent with our result, but the size of the emission region is one order of magnitude smaller than ours. Also, the magnetic field estimated in their model is much higher than what we found. The difference in some of the parameters could be because they modeled the average SED, and in our case, we are more focused on a short period of time. The optical-UV SED is mostly off to the general trend of broadband SED of blazar and hence in both cases, is fitted with a host-galaxy contribution. \citet{KKsingh_2019} used a specific model to fit the host-galaxy and estimated the black hole mass of the source, however, in JetSet we can not include a specific model, and hence host-galaxy is fitted as a free parameter.

\section{Conclusions}
\label{conclusion}
In this work, we present the long-term study of the blazar 1ES 1218+304 using 3.5 years of near-simultaneous multi-wavelength data from Fermi-LAT, SWIFT-XRT, SWIFT-UVOT, AstroSat, and TUBITAK observations taken between January 1, 2018, and March 31, 2021. This study explores the broadband temporal and spectral behavior of the source during flaring states. The main results of our study are provided below:
\begin{itemize}
    
    % Near simultaneous multi-wavelength light curves from UV to HE $\gamma$-ray have been reported here.
    \item The Astrosat SXT light curve reveals a variability of the order of 20 minutes and the X-ray spectrum is well fitted with both power-law and the log parabola models. However, the LP provides a better fit. A joint fit with the LAXPC spectrum provides a constraint on the location of synchrotron peak roughly around  $\sim$1.6 keV or 3.9$\times$10$^{17}$ Hz.
    \item The fast flux variability in $\gamma$-rays is calculated to be 0.39 days, the size of the emission region is estimated to be $\sim$2.3$\times$10$^{16}$ cm, and the emission region is located at a distance of $\sim 2.9 \times 10^{17}$ cm. A "harder-when-brighter" trend was seen in X-rays whereas a "softer-when-brighter" trend was in $\gamma$-rays. The $\gamma$-ray emission from 1ES 1218+304 can also be described by a power law with a spectral index of $\sim 1.75$.
    \item As seen in many other TeV blazars, a shift in synchrotron peak is observed from one state to another state from $\sim$10$^{17.5}$ Hz to possibly $\sim$10$^{20}$ Hz (from $\sim$1 keV to above 10 keV, possibly $\sim$500 KeV from our modeling) suggesting an extreme state of the source.
    \item The broadband SED modeling from radio to GeV energies is reproduced by a one-zone leptonic SSC model with the electron energy distribution described by a Power-law with an exponential cut-off (PLEC) function. We also find that the Optical/UV emissions from the source are dominated by the stellar thermal emissions from the host galaxy which are modeled by a simple blackbody approximation (\citealt{10.1111/j.1365-2966.2009.15738.x}) using JetSet. The JetSet code uses an approximation of the host galaxy model to help fit the SED modeling. We need more precise and dedicated observation in the UV/Optical band for a better understanding of the host galaxy.
\end{itemize}

\nocite{*}

\section*{Acknowledgements}
 We thank the referee for their insightful comments and suggestions which helped us to improve the draft.
D. Bose acknowledges the support of Ramanujan Fellowship-SB/S2/RJN-038/2017. 
R. Prince is grateful for the support of the Polish Funding Agency National Science Centre, project 2017/26/A/ST9/-00756 (MAESTRO 9) and the European Research Council (ERC) under the European Union’s Horizon 2020 research and innovation program (grant agreement No. [951549].
This work made use of Fermi telescope data and the Fermitool package obtained through the Fermi Science Support Center (FSSC) provided by NASA. This work also made use of publicly available packages JetSet, Fermipy, and PSRESP.
This publication uses the data from the AstroSat mission of the Indian Space Research Organisation (ISRO), archived at the Indian Space Science Data Centre (ISSDC). This work has used the data from the Soft X-ray Telescope (SXT) developed at TIFR, Mumbai, and the SXT POC at TIFR is thanked for verifying and releasing the data via the ISSDC data archive and providing the necessary software tools. We thank the LAXPC Payload Operation Center (POC) at TIFR, Mumbai for providing the necessary software tools.
We have also made use of the software provided by the High Energy Astrophysics Science Archive Research Center (HEASARC), which is a service of the Astrophysics Science Division at NASA/GSFC.

%%%%%%%%%%%%%%%%%%%%%%%%%%%%%%%%%%%%%%%%%%%%%%%%%%
\section*{Data Availability}
For this work, we have used data from the Fermi-LAT, Swift-XRT, Swift-UVOT, and AstroSat which are available in the public domain. We have also used optical data collected by the TUBITAK telescope. This optical data was given to us on request. Details are given in Section \ref{data_reduction}.



\bibliographystyle{mnras}
\bibliography{example} % if your bibtex file is called example.bib

%\begin{thebibliography}{99}
%\end{thebibliography}

%%%%%%%%%%%%%%%%%%%%%%%%%%%%%%%%%%%%%%%%%%%%%%%%%%

%%%%%%%%%%%%%%%%% APPENDICES %%%%%%%%%%%%%%%%%%%%%

%\appendix
%If you want to present additional material which would interrupt the flow of the main paper, it can be placed in an Appendix which appears after the list of references.

%%%%%%%%%%%%%%%%%%%%%%%%%%%%%%%%%%%%%%%%%%%%%%%%%%

% Don't change these lines
\bsp	% typesetting comment
\label{lastpage}
\end{document}

%% file: Tables/Astrosat_SXT.tex
%\begin{table}
%\begin{tabular}{|p{1.5cm}|p{1cm}|p{2cm}|}
%\begin{tabular}{ |c c c| }
%\hline
%\multicolumn{3}{|c|}{PowerLaw} \\
%\hline
%Model & Parameters & Value \\
%\hline
%TBabs & $N_H (10^{22} cm^{-2})$ & $0.0191$ \\
%Powerlaw & $\Gamma$ & $2.0025 \pm 0.0109227$ \\
% & Norm & $0.0261323 \pm 0.0001309$ \\
%Flux & $F_{0.3-7 keV}$ & $1.2205 \times 10^{-10}$ \\
% & $\chi^2/dof$ & $549.96/367$ \\
%\hline
%\end{tabular}
%\caption{Best fit spectral parameters of 1ES 1218+304 from SXT observations of 17-19 January 2019. X-ray flux is presented in unit (erg cm$^{-2}$ s$^{-1}$). $(\Delta \chi^2 = 1.498)$}
%\label{table:3}
%\end{table}

\begin{table*}
%\begin{tabular}{|p{1.5cm}|p{1cm}|p{2cm}|}
\caption {Best fit spectral parameters of 1ES 1218+304 from SXT observations of 17-20 January 2019. X-ray flux is presented in the unit (erg cm$^{-2}$ s$^{-1}$). The spectrum is fitted with both the power-law and log-parabola models. In the last row, we show the joint fit of the SXT and LAXPC spectrum. We also added a 3$\%$ systematic in the fit as suggested by the AstroSat team. The parameters are compared for free and fixed N$_H$ \citep{2016A&A...594A.116H} values. The overall fit provide better fit with free N$_H$.}
\begin{tabular}{ |c c c c c| }
%\hline
\multicolumn{3}{|c|}{} \\
\hline
Model & Parameters & Value \\

Power-law & &Fixed n$_H$ &  Free n$_H$ \\
\hline
TBabs & $N_H (10^{22} cm^{-2})$ & $0.0191$ & 0.057$\pm$0.005 \\
Index & $\Gamma$ & 1.95$\pm$0.01 & 2.11$\pm$0.02 \\
 %& Norm & $0.0261323 \pm 0.0001309$ \\
Flux & F$_{0.3-10.0}$ keV & $(1.427\pm0.004)\times 10^{-10}$ & $(1.474\pm0.006)\times 10^{-10}$ \\
 & $\chi^2/dof$ & $777/434$ & 595.75/433\\
\hline
Logparabola & & &   \\
\hline
TBabs & $N_H (10^{22} cm^{-2})$ & $0.0191$ & 0.11$\pm$0.02 \\
Index & $\alpha$ & 1.90$\pm$0.02 & 2.39$\pm$0.10 \\
      & $\beta$ & 0.28$\pm$0.04 & -0.38$\pm$0.13 \\
 %& Norm & $0.0261323 \pm 0.0001309$ \\
Flux & F$_{0.3-10.0}$ keV & $(1.300\pm0.009)\times 10^{-10}$ & $(1.81\pm0.05)\times 10^{-10}$ \\
 & $\chi^2/dof$ & $642.28/433$ & 590.55/432\\
\hline
Logparabola & joint fit & SXT + LAXPC &   \\
\hline
TBabs & $N_H (10^{22} cm^{-2})$ & $0.0191$ & 0.022$\pm$0.014 \\
Index & $\alpha$ & 1.88$\pm$0.02 & 1.89$\pm$0.07 \\
      & $\beta$ & 0.29$\pm$0.03 & 0.28$\pm$0.07 \\
 & Norm & 0.0264$\pm$0.0002 & $0.027 \pm 0.001$ \\
%Flux & F$_{0.3-10.0}$ keV & $(1.300\pm0.009)\times 10^{-10}$ & $(1.585\pm0.037)\times 10^{-10}$ \\
Constant factor& - & 0.95$\pm$0.04 & 0.95$\pm$0.04 \\ 
 & $\chi^2/dof$ & 556.19/392 & 556.08/391 \\
 \hline
\end{tabular}
\label{table:3}
\end{table*}

%% file: Tables/Fractional_variability.tex
\begin{table}
\centering
\begin{tabular}{|c|c|c|}
%\begin{tabular}{ |c c c| }
%\hline
%\multicolumn{3}{|c|}{PowerLaw} \\
\hline
Waveband & $F_{var}$ & err ($F_{var}$) \\
\hline
Fermi $\gamma$-ray & 0.2601 & 0.0964 \\
AstroSat-SXT X-ray & 0.0421 & 0.0058 \\
Swift X-ray & 0.5074 & 0.01513 \\
W1 & 0.9448 & 0.0006 \\
W2 & 0.6805 & 0.0005 \\
M2 & 0.9448 & 0.0007 \\
U & 0.0242 & 3.3185E-05 \\
V & 0.0147 & 0.0002 \\
B & 0.0171 & 0.0002 \\
R & 0.0144 & 6.5188E-05 \\
I & 0.0120 & 8.2755E-05 \\
\hline
\end{tabular}
\caption{
Fractional variability amplitude (F$_{var}$) parameter for the blazar 1ES 1218+304 from optical to HE $\gamma$-rays using observations during January 1, 2018 - May 31, 2021 (MJD 58119-59365) with different instruments.}
\label{table:6}
\end{table}

%% file: Tables/Fermi_spectrum_results.tex
\begin{table*}
\begin{tabular}{|c|c|c|c|c|c|c|c|}
%\begin{tabular}{ |c c c| }
%\hline
%\multicolumn{3}{|c|}{PowerLaw} \\

%\hline
%Spectral Index ($\alpha$) & -1.547 $\pm$ 0.230 & -1.540 $\pm$ 0.191 & -1.745 $\pm$ 0.030 & - \\
%Flux (F$_{0.3-300 GeV}$) & 3.306 & 3.063 & 1.310 & $10^{-8} \times $ photon(s) cm$^{-2}$ s$^{-1}$ \\
%Prefactor (N$_0$) & 9.538 $\pm$ 3.633 & 8.902 $\pm$ 2.796 & 2.966 $\pm$ 0.122 & $10^{-13} \times $ photon(s) cm$^{-2}$ s$^{-1}$ MeV$^{-1}$ \\
%TS & 43.497 & 48.297 & 2913.496 & - \\
%\hline

\hline
Parameter & Flare A & Flare B & Whole Time Period & Units \\
\hline
Spectral Index ($\alpha$) & -1.55 $\pm$ 0.23 & -1.54 $\pm$ 0.19 & -1.74 $\pm$ 0.03 & - \\
Flux (F$_{0.3-300 GeV}$) & 3.31 & 3.06 & 1.31 & $10^{-8} \times $ photon(s) cm$^{-2}$ s$^{-1}$ \\
Prefactor (N$_0$) & 9.54 $\pm$ 3.63 & 8.90 $\pm$ 2.80 & 2.97 $\pm$ 0.12 & $10^{-13} \times $ photon(s) cm$^{-2}$ s$^{-1}$ MeV$^{-1}$ \\
TS & 43.50 & 48.30 & 2913.50 & - \\
\hline
\end{tabular}
\caption{Best fit spectral parameters of 1ES 1218+304 from \texttt{Fermi}-LAT observations using equation \ref{eq:1} for two flaring periods 58488-58490 MJD (Flare A), 58498-58503 MJD (Flare B) and whole time period MJD 58119-59365.}
\label{table:5}
\end{table*}

%% file: Tables/Color_mag.tex
\begin{table*}
\begin{tabular}{|p{0.9cm}|p{1.5cm}|p{1.7cm}|p{1.3cm}|p{1.1cm}|p{1.3cm}|p{1.1cm}|}
%\begin{tabular}{|c|c|c|c|c|c|c|}
\hline
Colour Indices & Slope & Intercept & Pearson Coefficient & Pearson P-value & Spearman Coefficient & Spearman P-value \\
\hline
(B-V) & $0.216\pm0.024$ & $-3.152\pm0.390$ & 0.752 & 7.88E-13 & 0.774 & 6.33E-14 \\
(B-I) & $0.446\pm0.031$ & $-6.002\pm0.506$ & 0.893 & 1.15E-19 & 0.928 & 6.06E-24 \\
(R-I) & $0.156\pm0.019$ & $-1.982\pm0.317$ & 0.550 & 1.67E-05 & 0.734 & 2.65E-10 \\
(V-R) & $0.085\pm0.018$ & $-1.070\pm0.292$ & 0.745 & 1.52E-10 & 0.787 & 2.77E-12 \\
\hline
\end{tabular}
\caption{Colour magnitude fitting and correlations coefficient.}
\label{table:1}
\end{table*}

%\begin{table*}
%\begin{tabular}{|c{0.9cm}|c{1.8cm}|c{3.4cm}|}
%\begin{tabular}{|c|c|c|}
%\hline
%Band & C-Test & F-Test  \\
%& $C_1 , C_2, C$ & $F_1, F_2, F, F_c(0.99), F_c(0.999)$ \\
%%\hline
%B & 6.47, 6.84, 6.65 & 41.83, 46.80, 44.31, 1.38, 1.54 \\
%V & 5.95, 6.34, 6.14 & 35.35, 40.15, 37.75, 1.34, 1.48 \\
%R & 7.50, 7.66, 7.58 & 56.18, 58.62, 57.40, 1.33, 1.46 \\
%I & 6.13, 6.00, 6.06 & 37.54, 36.01, 36.77, 1.33, 1.46 \\
%\hline
%\end{tabular}
%\caption{Colour magnitude fitting and correlations coefficient.}
%\label{table:2}
%\end{table*}

%% file: Tables/SED_model.tex
\begin{table*}
%\begin{tabular}{|p{0.9cm}|p{1.5cm}|p{1.7cm}|p{1.3cm}|}
\begin{tabular}{|c||c|c|c|c|}
\hline

Sr. No. & Model Parameters & Unit & Flare A & Flare B \\
& &  & 5-7 Jan & 15-20 Jan \\
%\hline
%1. & $\gamma_{min}$ & - & $88.342$ & $5.9990$ \\
%2. & $\gamma_{max}$ & - & $6.3346\times10^{7}$ & $6.2115\times10^{7}$ \\
%3. & $\gamma_{cut}$ & - & $2.8153\times10^{7}$ & $6.2216\times10^{5}$ \\
%4. & R$_H$ & $10^{17}$cm & $1.0$ & 1.0 \\
%5. & R & $10^{16}$cm & $1.0658$ & 1.4 \\
%6. & $\alpha$ & - & $1.482500$ & $1.530156$ \\
%7. & N & cm$^{-3}$ & 85.34312 & 37.58231 \\
%8. & B & G & $2.7378\times10^{-3}$ & $1.3035\times10^{-2}$ \\
%9. & $z$ & - & $0.182$ & 0.182 \\
%10. & $\delta$ & - & $15.97827$ & $30.30340$ \\
%\hline
%11. & U$_e$ & erg cm$^{-3}$ & $3.470401$ & $4.179746\times10^{-2}$ \\
%12. & U$_B$ & erg cm$^{-3}$ & $2.982449\times10^{-7}$ & $6.760266\times10^{-6}$ \\
%13. & P$_e$ & erg s$^{-1}$ & $9.460334\times10^{45}$ & $7.081457\times10^{44}$ \\
%14. & P$_B$ & erg s$^{-1}$ & $8.130172\times10^{38}$ & $1.145346\times10^{41}$ \\
%15. & P$_{jet}$ & erg s$^{-1}$ & $1.060629\times10^{46}$ & $7.370064\times10^{44}$ \\
%\hline

\hline
1. & $\gamma_{min}$ & - & $88.34$ & $6.00$ \\
2. & $\gamma_{max}$ & - & $6.33\times10^{7}$ & $6.21\times10^{7}$ \\
3. & $\gamma_{cut}$ & - & $2.81\times10^{7}$ & $6.22\times10^{5}$ \\
4. & R$_H$ & $10^{17}$cm & $1.00$ & 1.00 \\
5. & R & $10^{16}$cm & $1.06$ & 1.40 \\
6. & $\alpha$ & - & $1.48$ & $1.53$ \\
7. & N & cm$^{-3}$ & 85.34 & 37.58 \\
8. & B & G & $2.74\times10^{-3}$ & $1.30\times10^{-2}$ \\
9. & $z$ & - & $0.18$ & 0.18 \\
10. & $\delta$ & - & $15.98$ & $30.30$ \\
\hline
11. & U$_e$ & erg cm$^{-3}$ & $3.47$ & $4.18\times10^{-2}$ \\
12. & U$_B$ & erg cm$^{-3}$ & $2.98\times10^{-7}$ & $6.76\times10^{-6}$ \\
13. & P$_e$ & erg s$^{-1}$ & $9.46\times10^{45}$ & $7.08\times10^{44}$ \\
14. & P$_B$ & erg s$^{-1}$ & $8.13\times10^{38}$ & $1.14\times10^{41}$ \\
15. & P$_{jet}$ & erg s$^{-1}$ & $1.06\times10^{46}$ & $7.37\times10^{44}$ \\
\hline

16. & Reduced Chi-Squared & - & 1.08 & 2.71 \\
\hline
\multicolumn{5}{|c|}{Host Galaxy} \\
\hline
17. & nuFnu$\_$p$\_$host & erg cm$^{-2}$ s$^{-1}$ & -10.37 & -10.37 \\
18. & nu$\_$scale & Hz & 0.50 & 0.49 \\
\hline
\end{tabular}
\caption{[1-3] Minimum, maximum and cut Lorentz factor of injected electron spectrum [4] The position of the region [5] The size of emission region [6] Spectral Index [7] Particle density [8] Magnetic field [9] Red Shift [10] Doppler factor [11] Electron energy density [12] Magnetic field energy density [13] Jet power in electrons [14] Jet power in magnetic field [15] Total jet power}
\label{table:4}
\end{table*}